\documentclass[letterpaper]{article} 
\usepackage[]{aaai2027}  
\usepackage[hyphens]{url}  
\usepackage{graphicx} 
\usepackage{natbib}  
\usepackage{caption} 
\usepackage{algorithm}
\usepackage{algorithmic}

\usepackage{newfloat}
\usepackage{listings}
\DeclareCaptionStyle{ruled}{labelfont=normalfont,labelsep=colon,strut=off} 
\floatstyle{ruled}
\newfloat{listing}{tb}{lst}{}
\floatname{listing}{Listing}
\usepackage{booktabs}
\usepackage{amsmath}
\usepackage{amssymb}
\usepackage{multirow}
\usepackage{enumitem}
\usepackage{framed}
\usepackage{dblfloatfix}

\lstdefinelanguage{json}{
  morestring=[b]",
  morecomment=[l]{//},
  morecomment=[s]{/*}{*/},
  sensitive=true
}

\newenvironment{promptbox}[1]{%
  \begin{framed}\noindent\textbf{#1}\par\medskip
}{\end{framed}} 
\title{ESG: Generating Physically Consistent Dynamic 3D Scenes from Text Descriptions}

\author{
    Xintong Fang\textsuperscript{\rm 1},
    Zhiyuan Fang\textsuperscript{\rm 2},
    Rengan Xie\textsuperscript{\rm 2},
    Xuhong Zhang\textsuperscript{\rm 1},
    Guoyuan An\textsuperscript{\rm 3},
    Zeran Liu\textsuperscript{\rm 1},
    Jingyan Zhang\textsuperscript{\rm 1},
    Jiarui Guo\textsuperscript{\rm 4},
    Yuchi Huo\textsuperscript{\rm 2}\thanks{Corresponding author.}
}
\affiliations{
    \textsuperscript{\rm 1}Zhejiang University, Hangzhou, China\\
    \textsuperscript{\rm 2}State Key Laboratory of CAD\&CG, Zhejiang University, Hangzhou, China\\
    \textsuperscript{\rm 3}Korea Advanced Institute of Science and Technology (KAIST)\\
    \textsuperscript{\rm 4}Northeastern University, China\\
    fxt525921@gmail.com, huo.yuchi.sc@gmail.com
}

\newcommand{\modi}[1]{#1}

\newcommand{\refFig}[1]{Figure \ref{#1}}

\newcommand{\refSup}[1]{Supp. Mat. \ref{#1}}
\newcommand{\refSec}[1]{Section \ref{#1}}
\newcommand{\refTab}[1]{Table \ref{#1}}

\newcommand{\Skip}[1]{}

\newcommand{\nn}{ESG}
\newcommand{\esg}{Evolutive Scene Graph}

\begin{document}

\maketitle

\begin{abstract}
Recent progress in image and 3D scene generation has enabled increasingly realistic static environments, yet most methods remain confined to such static configurations. Generating dynamic scenes from natural language is fundamentally challenging: it requires joint reasoning over scene structure, temporal evolution, and physical feasibility, while ensuring reliable execution in modern physics engines. We present a unified framework for generating physically consistent dynamic 3D scenes from text, with outputs directly executable in Unreal Engine. Central to our approach is the \emph{Evolutive Scene Graph} (ESG), which specifies entities with physical attributes, spatial relations, and event-driven timelines in a machine-checkable form. Given a prompt, a large language model constructs and validates a complete ESG; spatial layouts are grounded via energy-minimized gradient optimization; timeline-constrained physical parameters are then optimized through differentiable simulation to satisfy user-specified events; and the resulting scene is compiled into an engine-executable class. Experiments on 10 scenes across three complexity levels show that our method achieves $16.4/18$ mean event completion, outperforming
Scene Language, the strongest engine-executable baseline (SimWorld),
and our ablation without physical optimization by a clear margin in event completion and parameter accuracy.
\end{abstract}

\begin{figure}[t]
    \centering
    \includegraphics[width=\columnwidth]{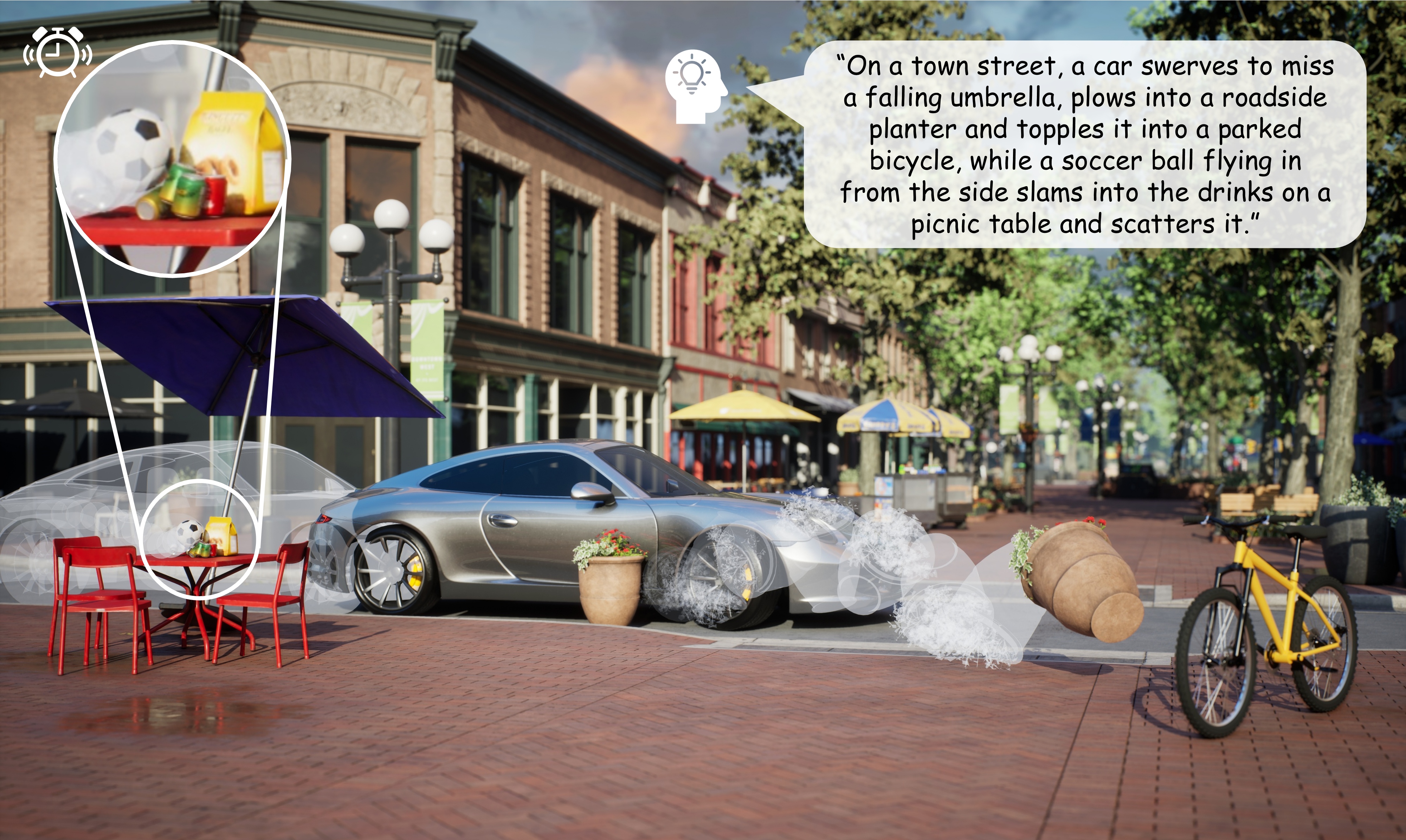}
    \caption{A high-quality dynamic 3D scene generated by our framework, where a textual description is translated into a physically consistent, executable Unreal Engine scene with multiple interacting objects and events.}
    \label{fig:teaser}
\end{figure}

\section{Introduction}

Recent advances in 3D generation have produced impressive results in
automatic scene synthesis, yet the majority of existing approaches
remain confined to static scenes, whereas many real-world
applications---films, games, virtual reality, and simulation---require
dynamic scenes in which objects interact and evolve under physical laws.

Dynamic scene generation introduces challenges well beyond static
layout synthesis: a scene must jointly model temporal structure,
event-driven interactions, and object-level physical parameters while
ensuring physically feasible motion. Deployment further demands direct
executability---explicit object instantiation, physics initialization,
and event logic compatible with real-time simulation frameworks.

Existing methods address only subsets of these requirements.
Learning-based approaches, including autoregressive and diffusion
models, predict static layouts or geometry distributions with no
temporal structure or physical dynamics~\cite{Paschalidou2021NEURIPS,tang2024diffuscene,zhai2023commonscenesgeneratingcommonsense3d,lin2024instructscene}.
Procedural systems such as Infinigen~\cite{raistrick2023infinigen}
scale to photorealistic worlds but offer no semantic or physical
controllability. SimWorld~\cite{ye2025simworld} and Virtual
Community~\cite{zhou2026virtualcommunity} build city-scale environments
atop real physics engines and deploy LLM agents for navigation, yet
translate agent commands into pre-defined motion primitives with no
mechanism to solve for the physical parameters that guarantee a
specified event---leaving completion dependent on agent policy rather
than physical optimality. PAT3D~\cite{lin2026pat3d} integrates
differentiable simulation into text-to-3D generation but targets static
equilibrium rather than temporal event satisfaction.

LLM- and script-based pipelines produce procedural or animation
descriptions from text, yet rely on heuristic parameters without
validating physical realizability~\cite{hu2024scenecraftllmagentsynthesizing,sun2025layoutvlmdifferentiableoptimization3d,yang2025sceneweaverallinone3dscene},
causing dynamic interactions to exhibit instability or inconsistency
in physics engines.

We argue that dynamic scene generation should be formulated as a
structured specification and physical grounding problem: natural
language is first translated into a machine-checkable representation
of entities, relations, events, and temporal constraints, then grounded
through physics-based optimization and simulation---enabling systematic
verification, controllability, and reliable execution that purely
generative approaches cannot guarantee.

Based on this insight, we propose a unified framework that translates
natural-language prompts into directly executable Unreal Engine scenes.
Central to our approach is the \emph{Evolutive Scene Graph} (ESG)---a
machine-checkable dynamic scene representation encoding physical
properties, semantic attributes, spatial relations, and event-driven
timelines in a standardized JSON schema. An LLM constructs and
validates the ESG through a layered process, resolving semantic
entities into concrete assets. Object poses are then grounded via
energy-minimized gradient optimization satisfying spatial constraints,
collision avoidance, and physical placeability.

The key technical contribution is timeline-constrained physical
optimization formulated as an event-constrained feasibility problem.
We optimize physical parameters---initial velocities, actuation forces,
timing offsets---by minimizing event-wise losses through differentiable
forward simulation, using an AD-default, FD-fallback gradient scheme
that balances efficiency with fidelity across interaction complexities.
The grounded ESG is compiled into UE physics initialization and event
scheduling logic.

Experiments demonstrate diverse dynamic scenes with stable physical
behaviors, high event completion rates, and low parameter errors across
a wide range of interaction types. We will publicly release our code.
Our contributions are threefold:

\begin{itemize}
  \item We introduce \emph{ESG (Evolutive Scene Graph)}, a structured
  representation that formally models objects, events, temporal
  constraints, and causal dependencies for 3D dynamic scenes, bridging
  natural-language descriptions and physically executable scene
  specifications.

  \item We formulate dynamic scene generation as an
  \emph{event-Constraint feasibility problem} and propose a hybrid
  optimization framework coupling LLM-driven motion template matching
  with timeline-aware, differentiable parameter search to satisfy
  multi-stage event constraints.

  \item We present a \emph{general differentiable physics framework}
  combining rigid-body dynamics with event-driven loss functions,
  enabling end-to-end optimization of physical parameters and direct
  compilation into an executable Unreal Engine environment.
\end{itemize}
\section{Related Work}

\subsection{3D Scene Generation}
\label{rw::sce}

Early 3D scene synthesis arranged retrieved assets via handcrafted
geometric rules~\cite{10.1145/2010324.1964981,10.1145/2366145.2366154};
PlanIT~\cite{10.1145/3306346.3322941} advanced this with hierarchical
grammars and interaction graphs, yet both remain confined to static
layouts.

Deep generative models shifted the paradigm to learning continuous
scene distributions. GAN- and VAE-based
methods~\cite{10.1145/3197517.3201362,ritchie2018fastflexibleindoorscene}
encode layouts into latent spaces; autoregressive models such as
ATISS~\cite{Paschalidou2021NEURIPS} treat synthesis as sequence
modeling; diffusion
models~\cite{zhai2023commonscenesgeneratingcommonsense3d,lin2024instructscene,tang2024diffuscene}
achieve superior scene coverage; and 3D-aware
generators~\cite{li2024director3d,devries2021unconstrainedscenegenerationlocally}
synthesize renderable NeRF scenes. SceneCraft~\cite{yang2024scenecraft}
scales to complex indoor topologies via semantic layouts and
diffusion-based rendering, but outputs a NeRF with no object
instantiation, physical properties, or dynamic control.

Broader-scale efforts enrich content but not dynamics.
Infinigen~\cite{raistrick2023infinigen} procedurally generates
photorealistic environments with no semantic or physical control.
City-scale platforms SimWorld~\cite{ye2025simworld} and Virtual
Community~\cite{zhou2026virtualcommunity} support LLM-driven
construction and agent navigation, but treat dynamics as agent
decisions rather than specifiable physical events, exposing no
mechanism to define or optimize timed object interactions.

LLMs have inspired language-driven scene planning.
GPT4Motion~\cite{lv2024gpt4motion} prompts GPT-4 to write Blender
physics scripts, but parameters are unoptimized and the output is
rendered video rather than simulation-ready data. More broadly,
LLM-based
planners~\cite{hu2024scenecraftllmagentsynthesizing,sun2025layoutvlmdifferentiableoptimization3d,yang2025sceneweaverallinone3dscene}
generate executable scripts yet validate feasibility only via
bounding-box heuristics, leaving event-level constraints unenforceable.

None of the above treats dynamic scene generation as event-constraint
satisfaction: outputs are static layouts or rendered videos with
feasibility checked only geometrically. We close this gap by
formulating scene generation as event-constrained parameter
optimization over a differentiable simulation loop, guaranteeing event
realization in a standard physics engine.

\subsection{Scene Description and Generation}
\label{rw::des}

Scene specifications bridge user intent and geometric instantiation. Dataset-driven approaches such as 3D-FRONT~\cite{fr20213d} inspired sequence-based scene models~\cite{wang2021sceneformerindoorscenegeneration}, while scene graphs~\cite{zhou2019scenegraphnet,graph2scene2021} capture richer inter-object relations; Graph-to-3D~\cite{graph2scene2021} decodes object attributes from semantic edges via graph convolutions.

LLMs have become direct reasoners over scene structure from natural language: LayoutGPT~\cite{feng2023layoutgpt} and InstructScene~\cite{lin2024instructscene} emit layout programs or semantic graphs for downstream generation; Holodeck~\cite{yang2024holodeck} converts spatial constraints into simulator-compatible scenes via constraint satisfaction; and Scene Language~\cite{zhang2025scenelanguage} represents scenes as programs, words, and embeddings for procedural 4D animation.

Existing specifications are nonetheless predominantly \emph{kinematic}: they describe static snapshots without temporal logic (e.g., ``object~$A$ hits object~$B$'') or physical properties (mass, friction), making them unsuitable for temporally-constrained interactions. Our evolutive scene graph addresses this by explicitly modeling event timelines and exposing object-level physical parameters.

\subsection{Physical Parameters Prediction and Optimization}
\label{rw::phy}

Observation-driven methods estimate physical parameters via differentiable simulation. Physics as Inverse Graphics~\cite{DBLP:journals/corr/abs-1905-11169} optimizes mass and friction to minimize trajectory error; NeuPhysics~\cite{qiao2022neuphysics} and PAC-NeRF~\cite{li2023pacnerfphysicsaugmentedcontinuum} embed differentiable dynamics into NeRF for joint geometry-physics identification—though all require visual supervision.

A parallel thread uses differentiable simulation for design and control. DiffTaichi~\cite{hu2019difftaichi} and DiffPD~\cite{du2021_diffpd} provide gradient-based substrates for rigid and deformable dynamics; PPR~\cite{yang2023ppr} and DSO~\cite{li2025dso} steer generative models toward stable configurations via simulation feedback. PAT3D~\cite{lin2026pat3d} couples vision-language models with a differentiable rigid-body simulator to optimize poses toward static equilibrium, but targets static stability rather than dynamic events.

WonderPlay~\cite{li2025wonderplay} and PhysGen3D~\cite{chen2025physgen3d} condition video diffusion on physical simulation for dynamic content generation, but optimize visual plausibility and produce rendered videos rather than simulation-ready parameters.

Our setting differs from all the above: parameters are inferred from natural language without visual supervision, optimized for dynamic event satisfaction across time, and compiled directly into an executable physics engine. LLM outputs serve as semantic physical priors refined through differentiable event-loss optimization.
\section{Method}

\begin{figure*}
\centering
\includegraphics[width=1\linewidth]{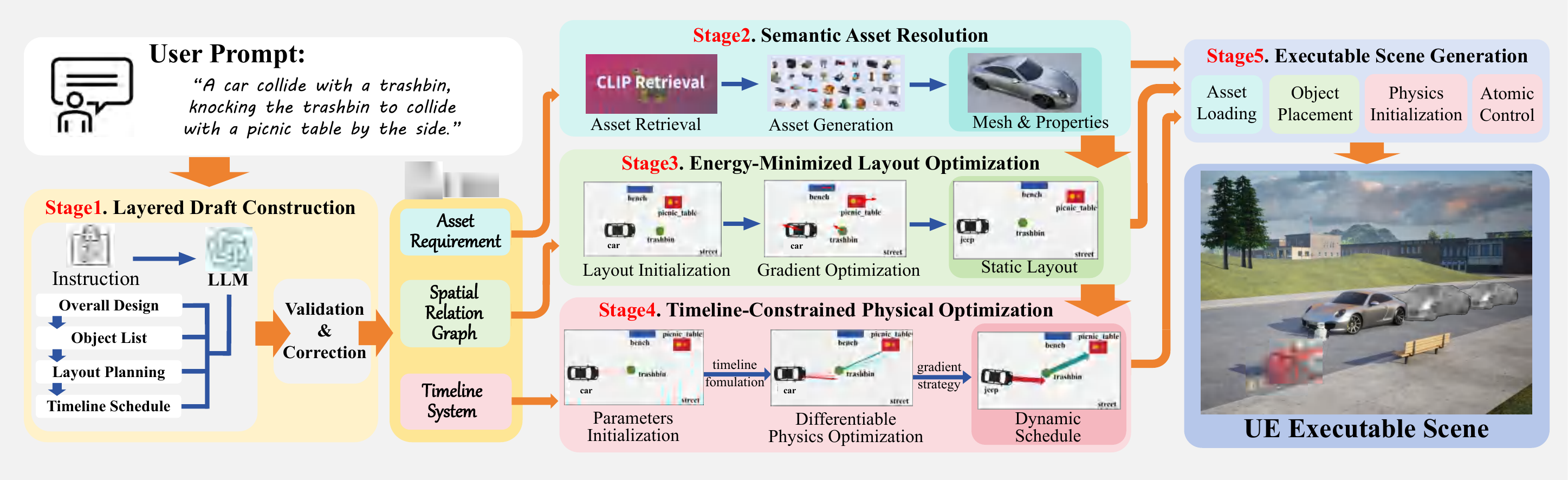}
\caption{\textbf{Method Overview.} Given a natural-language prompt, our system first constructs a machine-checkable \emph{\esg} (\nn) and resolves its semantic objects to simulation-ready assets. It then grounds the spatial specification through constraint-based layout optimization and grounds the event timeline through timeline-constrained physical optimization. The fully grounded \nn~is deterministically compiled into an executable Unreal Engine scene class.}
\label{method::pipeline}
\end{figure*}

\paragraph{Overview.}
As illustrated in~\refFig{method::pipeline}, our system generates a dynamic 3D scene in five stages, all operating on a single evolving carrier---the \emph{\esg~(\nn)}:
(1)~constructing a machine-checkable \nn~from the prompt under a standardized JSON schema;
(2)~resolving its semantic objects to simulation-ready assets;
(3)~grounding the spatial specification by optimizing object poses;
(4)~grounding the event timeline by optimizing control parameters through differentiable simulation;
and (5)~compiling the fully specified \nn~into an engine-executable scene class.

\subsection{Evolutive Scene Graph}

Directly mapping natural language to executable engine scripts is ill-posed: LLM outputs often violate the structural, causal, and numerical constraints required for deterministic physics simulation.
We therefore introduce \emph{\esg} (\nn), a machine-checkable intermediate representation that organizes user intent into a structured spatial specification and an executable event timeline.
\nn~also serves as the shared carrier of the pipeline: internal stages read the fields it requires and write back its results, progressively enriching one consistent, verifiable structure until compilable.

Given a prompt $q$, \nn~couples a spatial specification $\mathcal{G}$ with an event timeline $\mathcal{H}$:
\begin{equation}
\mathcal{G} = (O,E,R;\boldsymbol{\xi}), \qquad
\mathcal{H} = (P,I;\boldsymbol{\theta}),
\label{eq:esg-definition}
\end{equation}
where $O$, $E$, $R$ denote objects, typed spatial relations, and semantic regions. Each edge $e=(o_i{\to}o_j,\tau)\in E$ encodes a relation of type $\tau$; regions in $R$ provide coarse priors for layout initialization; and poses $\boldsymbol{\xi}=\{\boldsymbol{\xi}_i\}$ geometrically instantiate $\mathcal{G}$. $P$ and $I$ denote point and interval events, and $\boldsymbol{\theta}$ the control parameters realizing $\mathcal{H}$. Asset identities, $\boldsymbol{\xi}$, and $\boldsymbol{\theta}$ are initially unresolved and determined progressively by the subsequent stages.

\paragraph{Layered Construction of~\nn~(Stage~1).}
Given $q$, an LLM performs physics-aware expansion to infer implicit entities and interactions---e.g., introducing a staircase when an object rolls downward. A text-to-image model then renders a visual reference of the initial state, from which a VLM extracts soft priors over objects, relations, and regions. Conditioned on the prompt and these priors, the LLM instantiates $O$, $E$, $R$, $P$, $I$ under a type-constrained JSON schema, leaving $\boldsymbol{\xi}$ and $\boldsymbol{\theta}$ unresolved. A rule-based verifier checks schema validity, entity references, relation compatibility, and event ordering; violations are returned as structured feedback for iterative revision until all constraints hold.

\paragraph{Semantic Asset Resolution (Stage~2).}
Each object $o_i\in O$ is resolved to a simulation-ready asset: candidates are retrieved from an asset library via a CLIP-based~\cite{radford2021learningtransferablevisualmodels} embedding index~\cite{beaumont-2022-clip-retrieval}, then reranked by semantic similarity, category compatibility, and physical suitability. The selected asset is written back to \nn~with mesh path, canonical scale, collision proxy, and physics metadata. If retrieval confidence falls below a threshold, TRELLIS~\cite{xiang2024structured} generates the asset on demand; retrieved and generated assets share a unified schema, keeping downstream modules provenance-agnostic.

\paragraph{Progressive Physical Grounding (Stages~3--4).}
Stage~3 solves for poses $\boldsymbol{\xi}$ from asset-bound $\mathcal{G}$ via constraint-based layout optimization (\refSec{method::layout}); Stage~4 solves for $\boldsymbol{\theta}$ from the resulting layout and timeline $\mathcal{H}$ via differentiable physical optimization, ensuring every prescribed event is realized (\refSec{method::phyop}). Both write back into \nn.

\paragraph{Scene Compilation (Stage~5).}
The fully grounded ESG is finally compiled into an Unreal Engine scene class: event-level motion is mapped to physics initialization calls and event scheduling logic, emitting a C++ class that instantiates assets, sets initial physics states, and triggers events per the \nn~timeline. The class loads directly into a UE project, realizing both static layout and dynamic behaviors in a reproducible, editable form. Details are in~\refSup{sm::clsgen}.

\subsection{Energy-Minimized Layout Optimization}
\label{method::layout}

After the draft \nn~is grounded with engine assets, the object poses $\boldsymbol{\xi}$ remain unresolved and may induce unsatisfied spatial relations, geometric overlaps, or invalid support configurations. Holding $O$, $E$, and $R$ fixed, we determine $\boldsymbol{\xi}$ by minimizing a unified constraint energy, proceeding in two stages: constraint-aware initialization followed by multi-start gradient-based refinement.

\paragraph{Unified Constraint Energy.}
For the spatial specification $\mathcal{G}=(O,E,R;\boldsymbol{\xi})$, the relation energy over the spatial graph
\begin{equation}
\label{eq:relation-energy}
\Phi_{\mathrm{rel}}(\boldsymbol{\xi};E)
=
\sum_{e\in E}
w_e\,\Phi_{\tau}(\boldsymbol{\xi}_i,\boldsymbol{\xi}_j),
\end{equation}
where $\Phi_{\tau}$ measures the violation of relation type $\tau$ and $w_e=\alpha_{\tau}c_e$ weights each edge by its semantic strictness $\alpha_\tau$ and confidence $c_e$. The total layout energy incorporates collision and region constraints:
\begin{equation}
\label{eq:layout-energy}
\Phi
=
\Phi_{\mathrm{rel}}
+
\sum_{o_i\in O}
\left[
\lambda_{\mathrm{col}}
\Phi_{\mathrm{col}}^{i}(\boldsymbol{\xi})
+
\lambda_{\mathrm{reg}}
\Phi_{\mathrm{reg}}^{i}(\boldsymbol{\xi}_i,\rho_i)
\right],
\end{equation}
where $\Phi_{\mathrm{col}}^{i}$ and $\Phi_{\mathrm{reg}}^{i}$ measure, respectively, collisions involving $o_i$ and its violation of the assigned region $\rho_i\in R$. Pairwise collision penalties are distributed across incident objects so each collision is counted once. Together, these terms yield a differentiable objective balancing semantic consistency and physical placeability. We use a hybrid AABB--SDF collision formulation; definitions are in~\refSup{sm::layopt}.

\paragraph{Constraint-Aware Layout Initialization.}
Minimizing Eq.~\ref{eq:layout-energy} from arbitrary poses often leads to severe initial collisions and poor local minima. We extract placement dependencies from $E$, organize them as a DAG, and obtain a topological order $\pi=(\pi_1,\ldots,\pi_{|O|})$. Objects are then placed sequentially: let $\Omega_i\subseteq\rho(o_i)$ be the feasible pose domain for $o_i$ after accounting for region boundaries and asset geometry. For the $k$-th initialization, we sample a candidate set $\mathcal{C}_{\pi_t}^{k}\subset\Omega_{\pi_t}$ and select
\begin{equation}
\label{eq:layout-initialization}
\boldsymbol{\xi}_{\pi_t}^{k,0}
=
\arg\min_{\boldsymbol{\xi}\in\mathcal{C}_{\pi_t}^{k}}
\Delta\Phi
\!\left(
\boldsymbol{\xi}
\,\middle|\,
\boldsymbol{\xi}_{\pi_{<t}}^{k,0}
\right),
\end{equation}
where $\Delta\Phi$ is the energy increase from inserting the candidate into the current partial layout, comprising the active relation, collision, and region terms from Eq.~\ref{eq:layout-energy}. Initialization and refinement thus optimize the same objective.

\paragraph{Multi-start Gradient-based Optimization.}
Since $\Phi$ couples multiple spatial and geometric constraints, its landscape is highly non-convex. Repeating the sequential sampling with independent candidate sets and region-conditioned proposals yields $K$ diverse initializations $\{\boldsymbol{\xi}^{k,0}\}_{k=1}^{K}$. All poses are then jointly refined from each:
\begin{equation}
\label{eq:multi-start-optimization}
\widehat{\boldsymbol{\xi}}^{k}
=
\operatorname{GD}(\Phi,\boldsymbol{\xi}^{k,0}),
\qquad
\boldsymbol{\xi}^{*}
=
\arg\min_{\boldsymbol{\xi}\in
\{\widehat{\boldsymbol{\xi}}^{k}\}_{k=1}^{K}}
\Phi,
\end{equation}
where $\operatorname{GD}(\Phi,\boldsymbol{\xi}^{k,0})$ denotes gradient descent initialized at $\boldsymbol{\xi}^{k,0}$, terminating when the relative energy change falls below $\epsilon$ or the iteration limit is reached. The lowest-energy result $\boldsymbol{\xi}^{*}$ instantiates the final spatial specification $\mathcal{G}$.
\subsection{Timeline-Constrained Physical Optimization}
\label{method::phyop}

Our dynamic stage solves for physically valid motion under an explicit timeline, rather than relying on keyframes or spline overrides that can violate dynamical consistency. By formulating dynamic scene generation as an event-constrained feasibility problem, we avoid over-specifying motion details typically absent from natural-language descriptions.

\paragraph{Event Formulation.}
Recalling the event timeline $\mathcal{H}=(P,I;\boldsymbol{\theta})$, each event $h\in P\cup I$ identifies participating objects from $O$, partitioned into active bodies governed by $\boldsymbol{\theta}$ and passive bodies that evolve through physics, with remaining objects serving as static geometry. A point event $p\in P$ specifies a discrete condition (e.g., \textsc{Collision}, \textsc{PositionAt}) realized at a fixed time $t_p$ or within a window $[t_p^{\min},t_p^{\max}]$. An interval event $i=(u,v)\in I$ prescribes a motion regime (e.g., \textsc{UniformMotion}, \textsc{Brake}) between two bounding point events, with optional constraints on speed or heading. This formulation decouples \emph{what} must happen at discrete events from \emph{how} motion evolves between them, enabling event-level supervision while leaving continuous trajectories to physics.

\paragraph{Optimization Objective.}
Each event contributes a task loss $\mathcal{L}_{\mathrm{task}}^{h}$ encoding the required physical outcome and a temporal loss $\mathcal{L}_{\mathrm{win}}^{h}$ penalizing realization outside its admissible window. A global prior $\mathcal{L}_{\mathrm{p}}$ prevents parameters from drifting to implausible values. The total objective is
\begin{equation}
\label{eq:total-loss}
\mathcal{L}(\boldsymbol{\theta})
=
\sum_{h\in P\cup I}
\!\left(
w_{\mathrm{task}}^{h}\mathcal{L}_{\mathrm{task}}^{h}
+
w_{\mathrm{win}}^{h}\mathcal{L}_{\mathrm{win}}^{h}
\right)
+
w_{\mathrm{p}}\mathcal{L}_{\mathrm{p}}.
\end{equation}

For the dominant \textsc{Collision} event between bodies $A$ and $B$, let $\mathcal{S}$ be the simulated rollout frames. We identify the closest-approach frame
$\hat{s}=\arg\min_{s\in\mathcal{S}} d_{\mathrm{surf}}(A(s),B(s))$
and set $\hat{t}=\hat{s}\Delta t$. The losses are
\begin{equation}
\label{eq:collision-loss}
\begin{aligned}
\mathcal{L}_{\mathrm{task}}^{\textsc{col}}
&=\bigl[d_{\mathrm{surf}}(A(\hat{s}),B(\hat{s}))\bigr]_{+}^{2},\\
\mathcal{L}_{\mathrm{win}}^{\textsc{col}}
&=[t_{\mathrm{lo}}-\hat{t}]_{+}^{2}+[\hat{t}-t_{\mathrm{hi}}]_{+}^{2},
\end{aligned}
\end{equation}
where $[x]_{+}=\max(x,0)$ and $d_{\mathrm{surf}}$ is the signed surface-to-surface distance from the analytic proxies established during layout optimization. The task loss vanishes at contact; the temporal loss vanishes when contact occurs within $[t_{\mathrm{lo}},t_{\mathrm{hi}}]$; a fixed-time event sets $t_{\mathrm{lo}}=t_{\mathrm{hi}}$. The prior $\mathcal{L}_{\mathrm{p}}$ is a normalized $\ell_2$ penalty relative to $\boldsymbol{\theta}^{0}$. All losses are normalized by type-specific reference scales before summation; remaining event types are detailed in~\refSup{sm::phyopdt}.

\paragraph{Gradient Computation.}
We evaluate event losses using Newton, the rigid-body simulation layer of NVIDIA Warp~\cite{warp2022}, whose accurate contact resolution ensures the objective reflects true physical outcomes. For efficient gradient computation, we construct a differentiable surrogate that integrates gravity and damping analytically and models body--body interactions via soft-contact force $\mathbf{f}=k_e[-d]_{+}\hat{\mathbf{n}}$, enabling automatic differentiation (AD) through the full rollout in a single forward--backward pass.

For direct interactions, AD gradients are reliable and serve as the default. For chain interactions, where upstream collisions alter downstream trajectories, the simplified contact model may misrepresent post-collision momentum transfer; we therefore fall back to finite-difference (FD) gradients on the full Newton simulator at the cost of $2|\boldsymbol{\theta}|$ forward simulations per step. This AD-default, FD-fallback scheme balances gradient efficiency with physical fidelity. The optimized parameters $\boldsymbol{\theta}^{*}$ are written back to $\mathcal{H}$, completing the fully grounded dynamic scene specification.
\section{Experiment}

\begin{figure*}[t]
  \begin{minipage}[t]{0.48\linewidth}
    \centering
    \includegraphics[width=\linewidth]{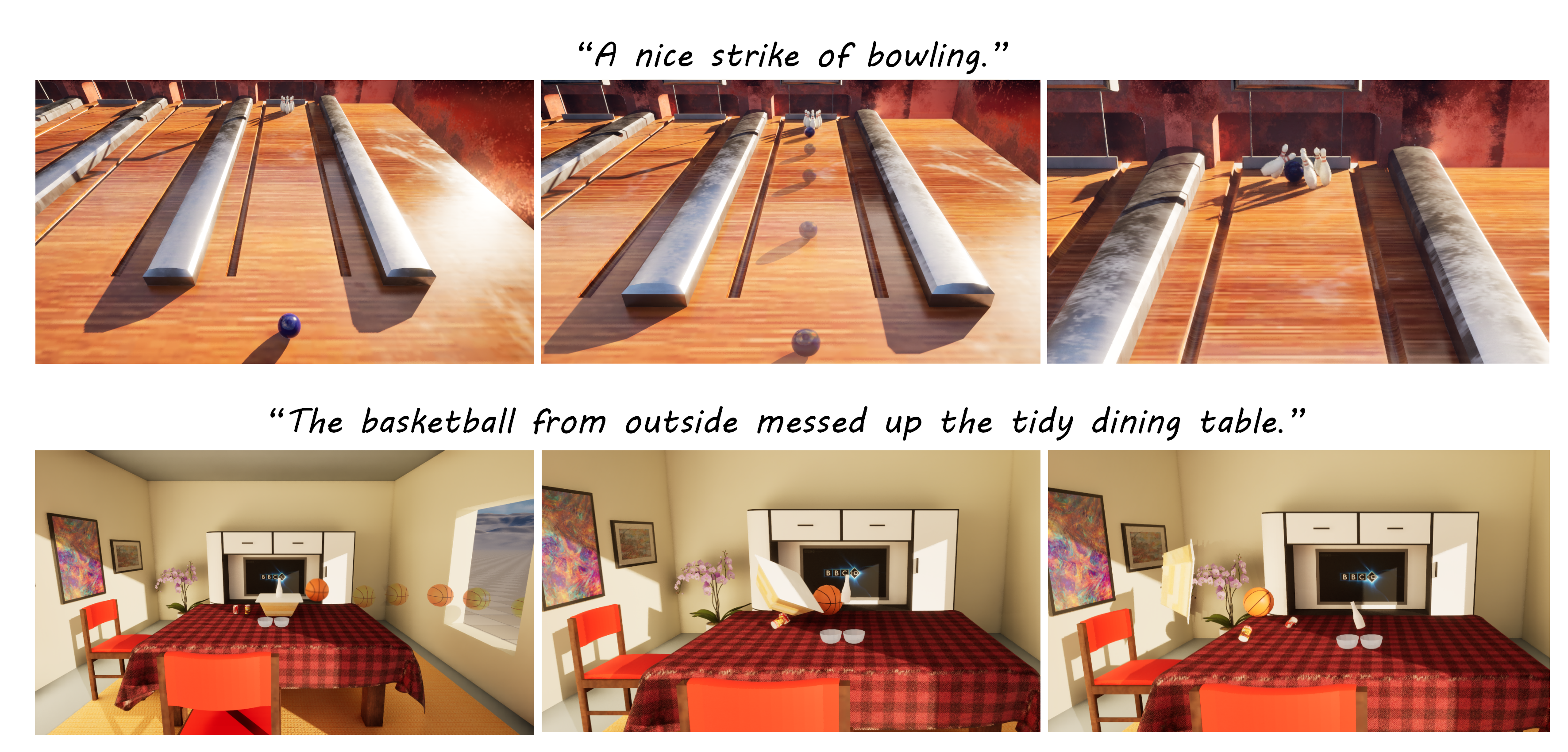}
    \captionof{figure}{Scene generation from simple input. The system infers and places implicit scene elements (e.g., bowling alley, tableware) not explicitly mentioned, producing physically plausible, semantically coherent layouts.}
    \label{experiment::simple_input}
  \end{minipage}
  \hfill
  \begin{minipage}[t]{0.48\linewidth}
    \centering
    \includegraphics[width=\linewidth]{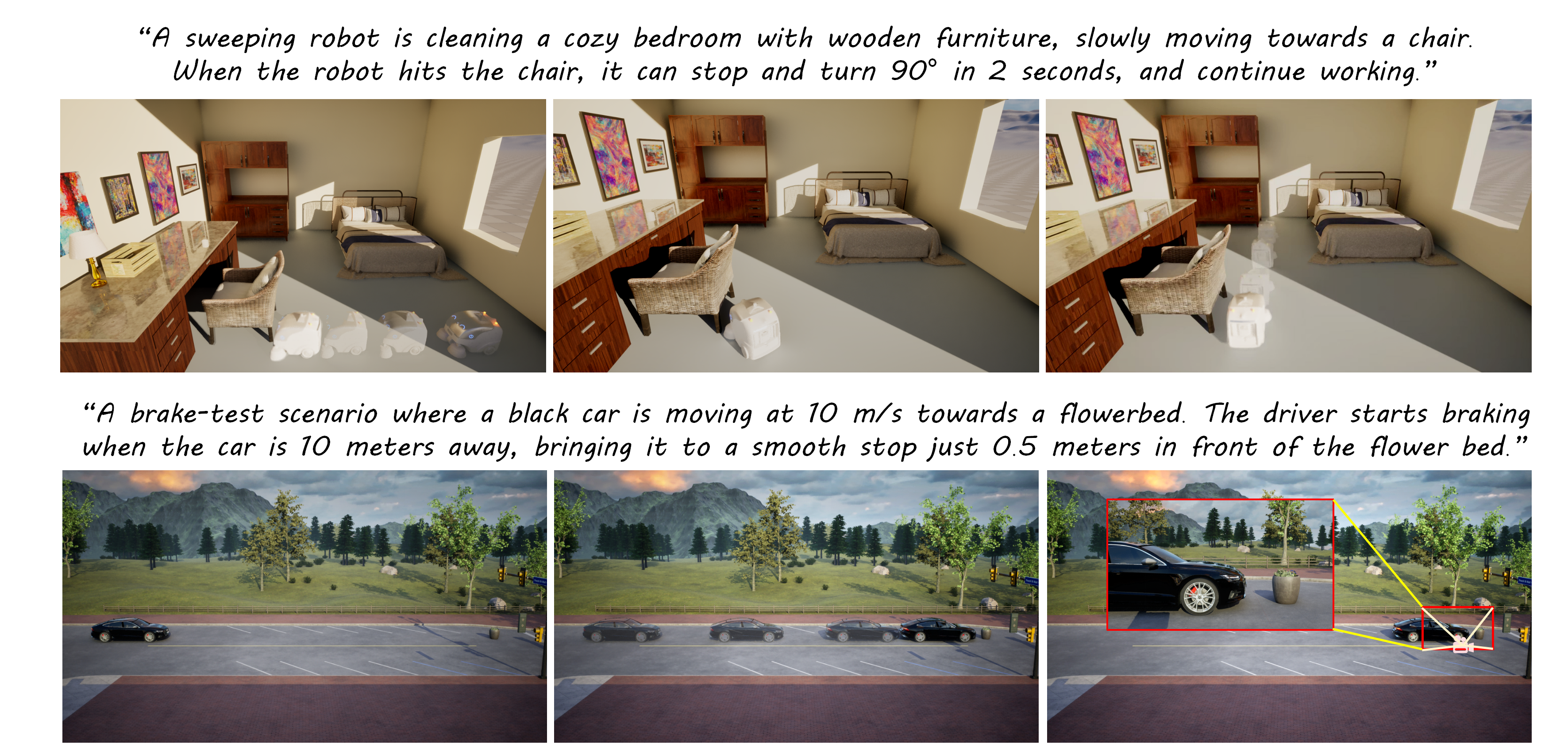}
    \captionof{figure}{Scene generation from complex input. Explicitly specified parameters (speed, turning angle, etc.) are treated as hard constraints; remaining parameters are optimized within the feasible solution space.}
    \label{experiment::complex_input}
  \end{minipage}
\end{figure*}

\begin{figure*}[t]
  \begin{minipage}[t]{0.48\linewidth}
    \centering
    \includegraphics[width=\linewidth]{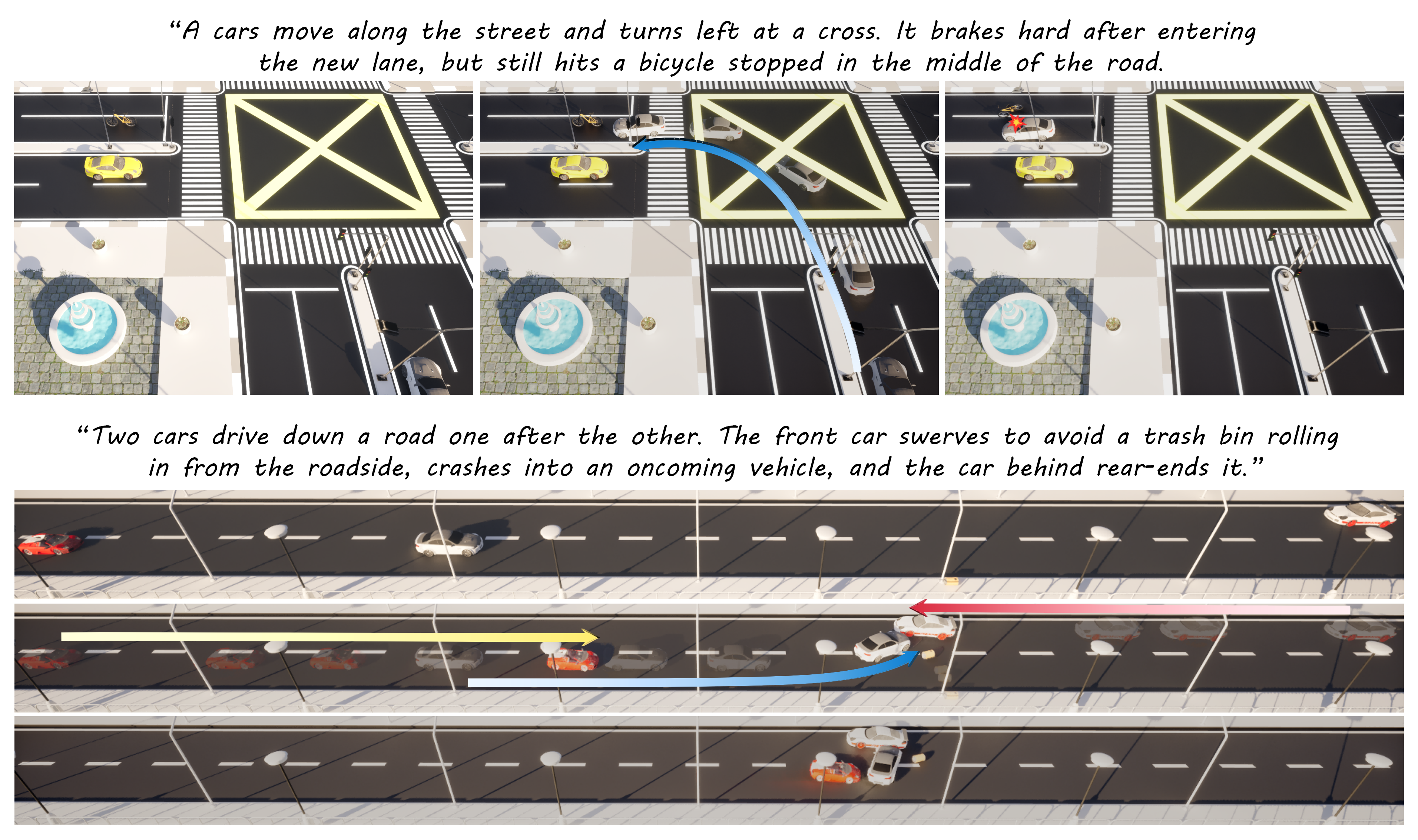}
    \captionof{figure}{Long dynamic sequence generation. The timeline representation supports long-horizon, multi-event specifications with parallel actions, enabling simultaneous control of multiple vehicles through extended maneuver sequences.}
    \label{experiment::long_sequence}
  \end{minipage}
  \hfill
  \begin{minipage}[t]{0.48\linewidth}
    \centering
    \includegraphics[width=\linewidth]{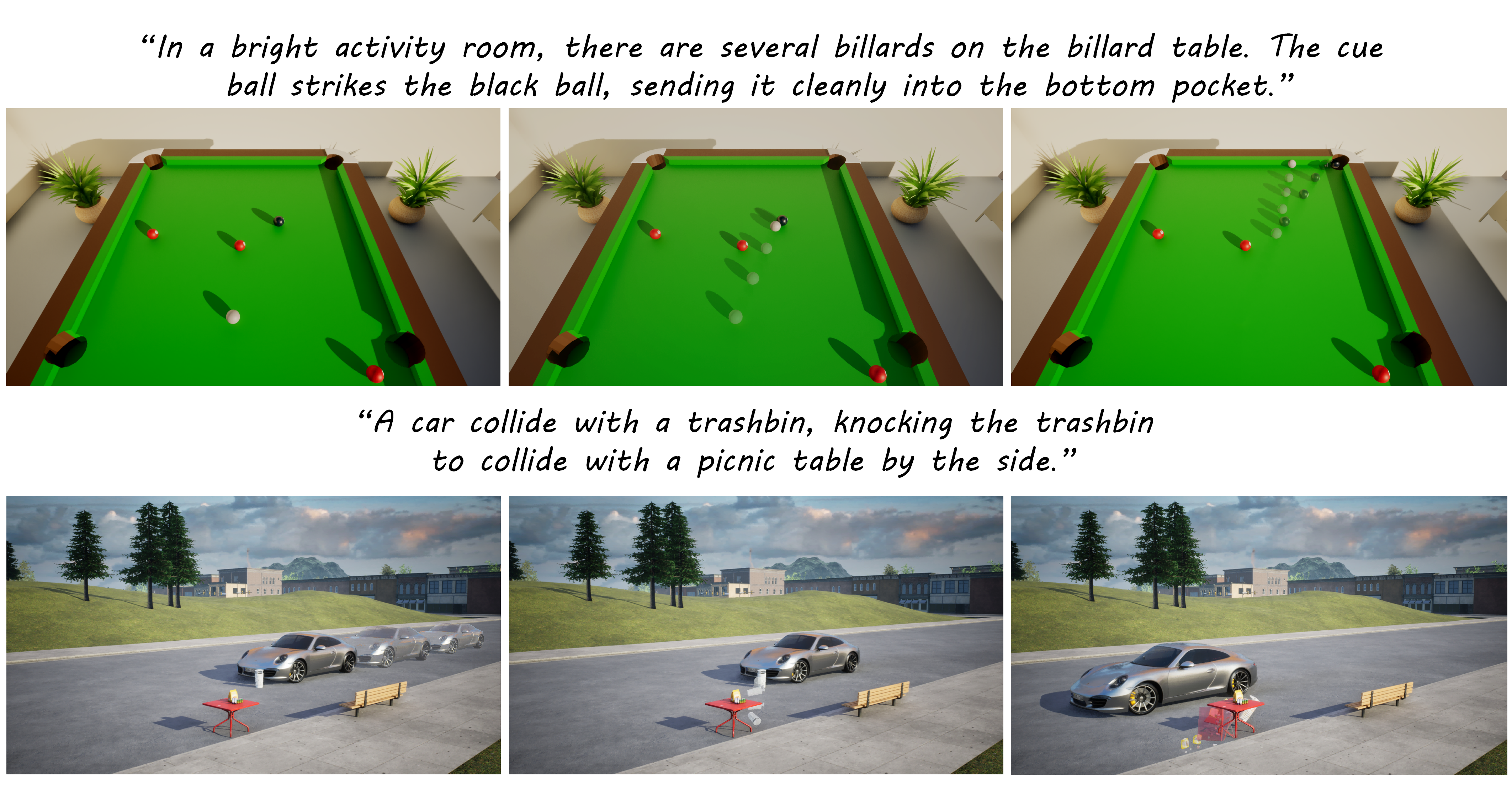}
    \captionof{figure}{Multi-stage cascading motion. A multi-stage event loss is backpropagated through the full causal chain (e.g., cue ball $\to$ black ball $\to$ pocket), enabling precise control of arbitrary chain reactions.}
    \label{experiment::cascade}
  \end{minipage}
\end{figure*}

\subsection{Quality Assessment}
We evaluate the system on a representative suite of prompts covering diverse requirements. ~\refFig{experiment::simple_input} and~\refFig{experiment::complex_input} test robustness to prompt specificity: simple inputs demonstrate plausible completion of missing elements while preserving narrative coherence; complex inputs with explicit numeric constraints, asset cues, and temporal requirements are followed with high fidelity and no spurious extrapolations.

For more complex scenarios, ~\refFig{experiment::long_sequence} and~\refFig{experiment::cascade} show long-sequence operations and multi-stage cascading interactions. Results remain aligned with the narrative throughout; downstream motions are refined by backpropagating event losses from upstream interactions, yielding controllable parameter updates. More experiments in ~\refSup{sm::assessment}.

\begin{figure*}[!htbp]
  \centering
  \includegraphics[width=\linewidth]{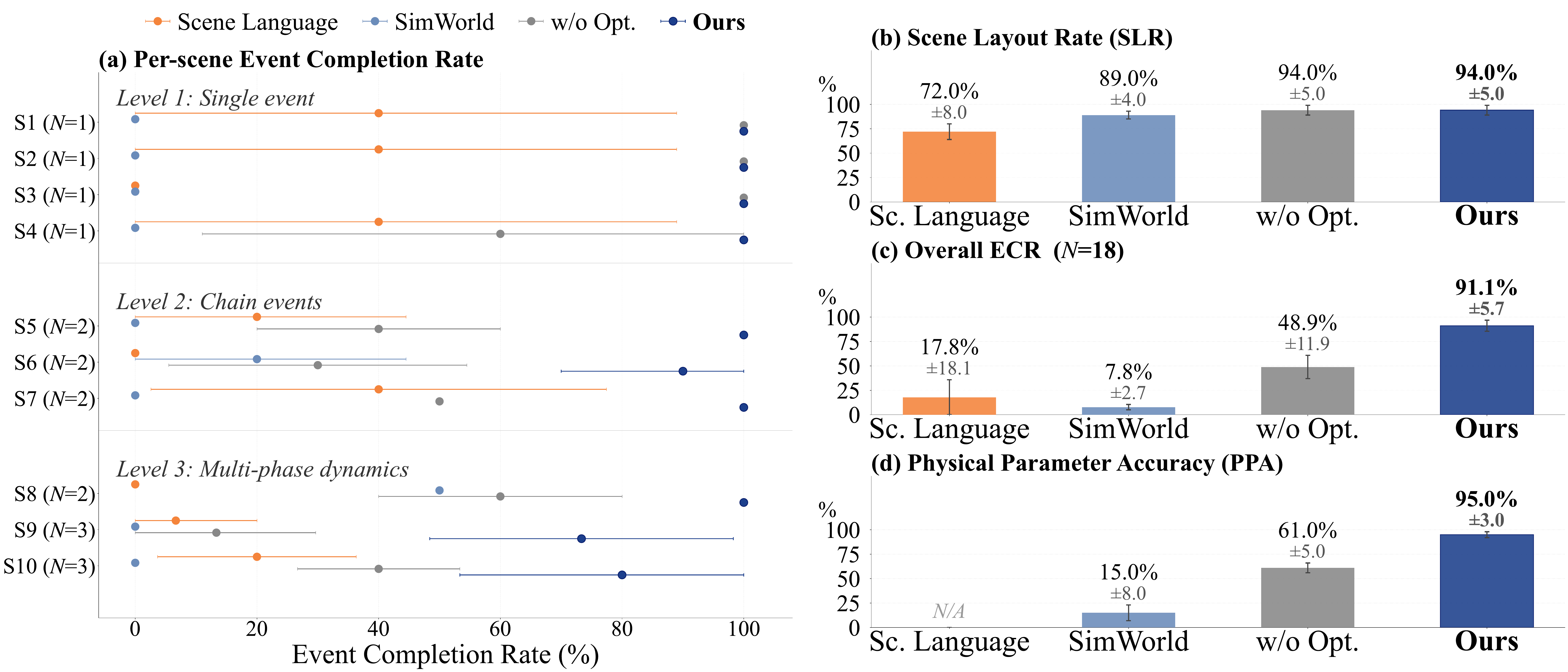}
  \caption{Comparison on 10 test scenes ($K{=}5$ runs, mean\,$\pm$\,1\,std).
    \textbf{(a)} Per-scene ECR by complexity level.
    \textbf{(b--d)} Overall SLR, ECR, and PPA across methods;
    \textbf{w/o Opt.}\ denotes our method without differentiable
    optimization.}
  \label{experiment::results}
\end{figure*}

\begin{figure*}[!htbp]
  \begin{minipage}[t]{0.48\linewidth}
    \centering
    \includegraphics[width=\linewidth]{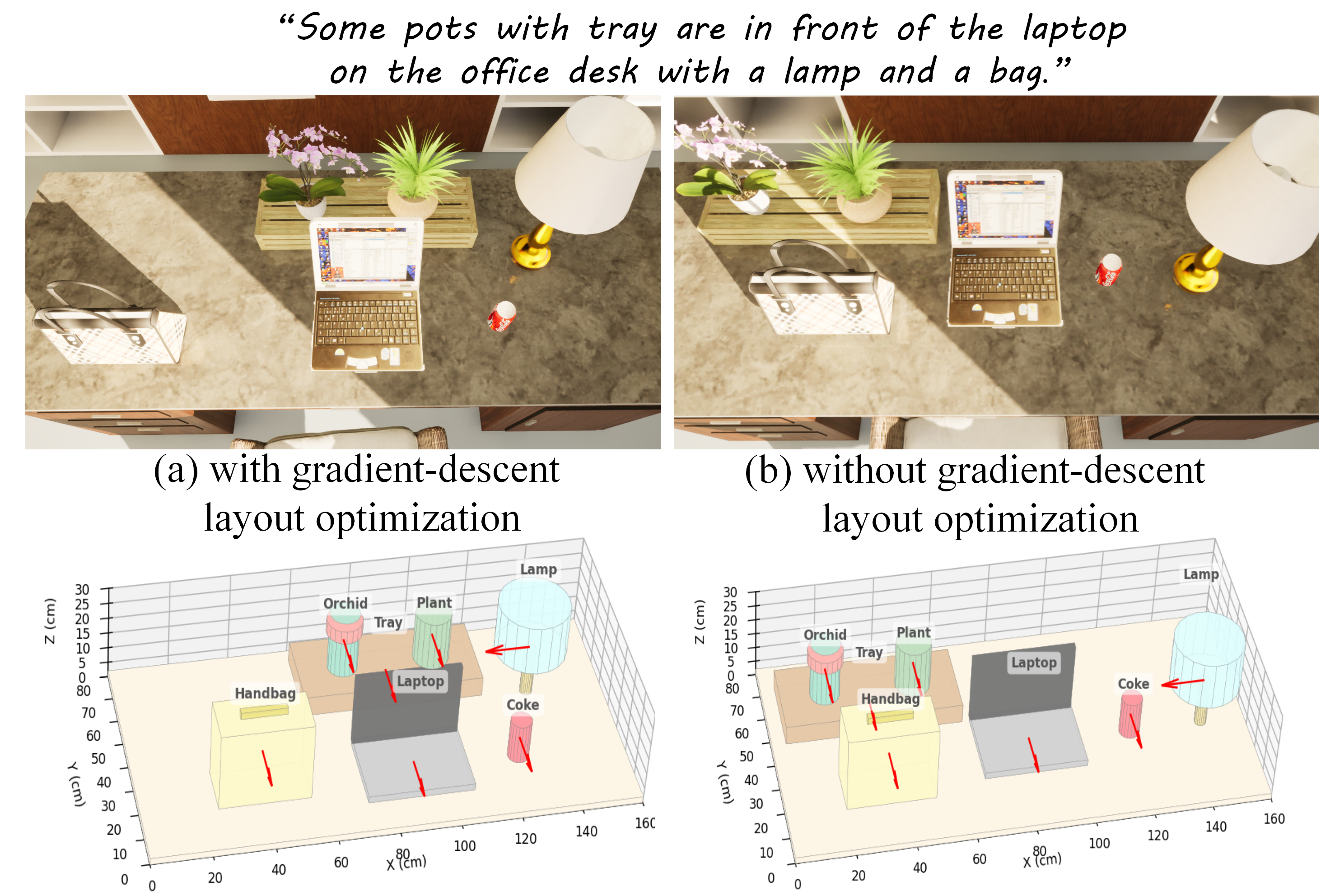}
    \captionof{figure}{Ablation of global gradient-descent optimization. Laptop captures center position, leaving no space for the tray.}
    \label{experiment::Ablation1}
  \end{minipage}
  \hfill
  \begin{minipage}[t]{0.48\linewidth}
    \centering
    \includegraphics[width=\linewidth]{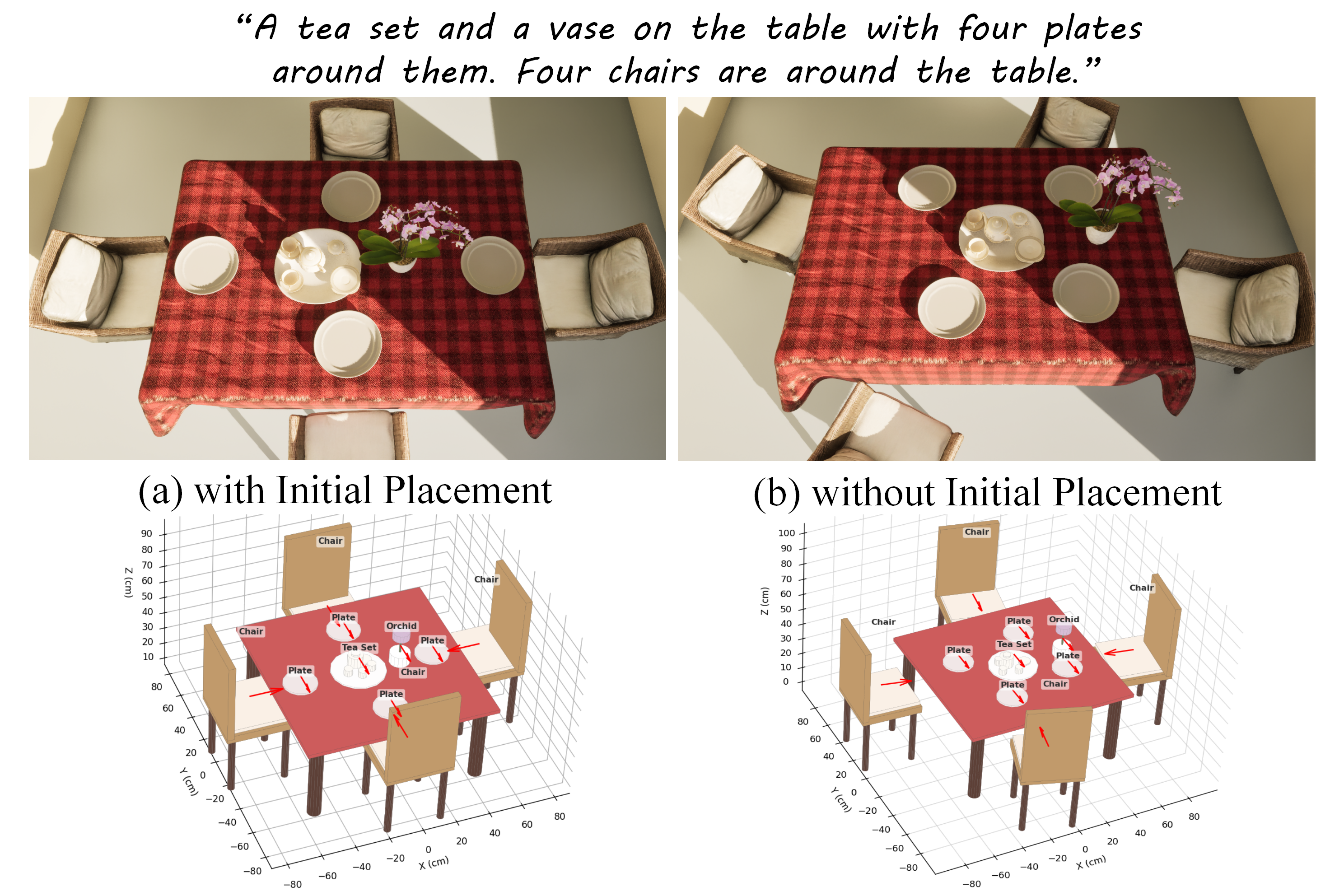}
    \captionof{figure}{Ablation of initial placement. The preference-based strategy produces a more orderly starting layout.}
    \label{experiment::Ablation2}
  \end{minipage}
\end{figure*}

\subsection{Comparison with baselines}
No public benchmark exists for text-to-dynamic scenes with
engine-verifiable events.
We therefore construct 10 scenes that progressively stress physical
feasibility and interaction complexity---comparable in scale to Scene
Language~\cite{zhang2025scenelanguage} and PAT3D~\cite{lin2026pat3d}
(17/18 prompts)---at three levels:
\emph{single-event} (S1--S4);
\emph{chain events} (S5--S7), requiring multi-step momentum transfer;
and \emph{multi-phase dynamics} (S8--S10), with concurrent or
time-windowed phases.
The suite has 18 timed events, matching these prompt suites in
evaluation load while covering longer causal chains under rigid-body
physics.
Full definitions are in~\refSup{sm::comparison}.

\paragraph{Evaluation metrics.}
We adopt three engine-native metrics motivated by simulation-grounded
benchmarks~\cite{li2025fate,worldarena2025}.
\textbf{SLR} checks static validity: all objects spawn with penetration
$\delta{<}2$\,cm.
\textbf{ECR} counts an event complete when native callbacks confirm
sufficient impulse, bounded penetration, and a $+0.2$\,s post-hit recheck
rules out tunneling; chain events require ordered completion.
\textbf{PPA} assigns each prompt-specified parameter a continuous score
in $[0,1]$: satisfied constraints score~$1$; violations incur a
proportional penalty scaled by constraint magnitude (interval half-width
or point target).
Initial-condition parameters are scored over all $K{=}5$ runs; outcome
parameters score non-completion as~$0$.

\paragraph{Baselines.}
Video-generation methods (GPT4Motion~\cite{lv2024gpt4motion},
WonderPlay~\cite{li2025wonderplay}, PhysGen3D~\cite{chen2025physgen3d})
target perceptual video synthesis rather than parametric scenes with
explicit physical states, lacking object-level position, velocity, or
collision records---incommensurable with our task by design.
Static-layout generators (Infinigen~\cite{raistrick2023infinigen},
LayoutGPT~\cite{feng2023layoutgpt}, Holodeck~\cite{yang2024holodeck})
produce no dynamic output. SceneCraft~\cite{hu2024scenecraftllmagentsynthesizing}
and PAT3D~\cite{lin2026pat3d} score 0\%~ECR, confirming spatial
arrangement alone cannot realize dynamic events.
We compare Scene Language~\cite{zhang2025scenelanguage}, imported into UE
by extracting poses and inferring launch velocities from kinematic trails
(Supp.~Mat.~\ref{supp:sl-adapter});
SimWorld~\cite{ye2025simworld}, whose dynamics follow agent policy rather
than physical-parameter optimization; and \textbf{w/o Opt.}, our pipeline
without gradient refinement. All methods share prompts and ECR
instrumentation ($K{=}5$).

\paragraph{Experiment Results.}
\refFig{experiment::results} reports all three metrics.
Scene Language achieves $\mathbf{3.2/18}$ ECR (SLR\,$\approx$\,72\%):
occasional single-collision successes but no causal mechanism for chain
events; high variance reflects its stochasticity, and PPA is
inapplicable.
SimWorld achieves $\mathbf{1.4/18}$ (SLR\,$\approx$\,89\%): fixed
animations yield large parameter errors ($|\Delta t|{=}0.50$\,s,
$|\Delta v|{=}1.96$\,m/s) and PPA of only $15.0\%$.
\textbf{w/o Opt.} achieves $\mathbf{8.8/18}$ (SLR\,$\approx$\,94\%):
simple collisions succeed (S1--S3: $1.0_{\pm0.0}$), but second-stage
chain contacts and complex maneuvers fail without gradient feedback.
Overall PPA is $61.0\%$ (IC\,$\approx$\,$100\%$;
outcome\,$\approx$\,$37\%$).
Our method achieves $\mathbf{16.4/18}$ (SLR\,$\approx$\,94\%) across all
levels, with PPA $95.0\%$ (IC\,$\approx$\,$100\%$,
outcome\,$\approx$\,$92\%$; $|\Delta t|{=}0.07$\,s,
$|\Delta v|{=}0.23$\,m/s, $|\Delta m|{=}0.02$\,kg).
The 7.6-point ECR gap and 55-point outcome PPA gap ($37\%$ vs.\ $92\%$)
show that differentiable event-loss optimization---not LLM
initialization---is decisive.

\subsection{Ablation Study}
We ablate four components: layout optimization, physical parameter
optimization, draft construction, and asset generation. We further
ablate the gradient strategy (AD-default, FD-fallback) and the
collision model (proxy with SDF refinement). Human evaluation and additional ablations are
in~\refSup{sm::human} and~\refSup{sm::MoreAb}.

\paragraph{Ablation Study of ESG Draft Construction.}
\label{exp::AbDraft}
We ablate hierarchical prompting (progressive staged generation) and
downstream validation (schema correctness and contradiction detection).
Results in~\refTab{tab:ablation_draft} confirm both contribute
substantially; details in~\refSup{sm::AbDraft}.

\begin{table}[!htbp]
\centering
\caption{ESG draft construction ablation.
Usability: percentage of error-free executable drafts.
High-Quality $|$ Usable: percentage of usable drafts admitting
low-energy layout.}
\label{tab:ablation_draft}
\resizebox{\columnwidth}{!}{
\setlength{\tabcolsep}{7pt}
\begin{tabular}{lcc}
\toprule
\textbf{Method}
  & \textbf{Usability (\%)}
  & \textbf{High-Quality $|$ Usable (\%)} \\
\midrule
Single-shot prompt      & 76.5 & 58.0 \\
w/o Validation          & 88.5 & 79.7 \\
\textbf{Ours (Hierarchical + Validation)} & \textbf{99.0} & \textbf{89.5} \\
\bottomrule
\end{tabular}
}
\end{table}

\paragraph{Ablation Study of Layout Optimization.}
As shown in~\refFig{experiment::Ablation1}, greedy one-off placement
evaluates poses only against already-placed objects and cannot revise
them, causing dead-ends: the laptop occupies the center, leaving no
feasible spot for the tray. Without global refinement, soft relations
(\textit{near}) cannot be jointly satisfied and penetrations remain.
As shown in~\refFig{experiment::Ablation2}, initializing all objects
at region centers yields symmetric configurations; the optimizer
relies on noise to break symmetry, producing penetration-free but
spatially underdetermined layouts.
Both confirm that reliable layouts need sampling-driven initialization
with global gradient-based refinement.

\paragraph{Ablation Study of Gradient Strategy.}
We compare three strategies on scene S6 ($K{=}5$);
see~\refTab{tab:ablation-grad}.
AD-only completes all first-stage events (5/5) but fails on the chain
event (0/5): backpropagating through non-spherical car--trashbin
contact yields directional errors in surrogate normals.
FD-only achieves 90\% via exact Newton physics at $4.4{\times}$
runtime. Our Hybrid matches FD accuracy at 90\% while cutting time by
33\%, using AD for fast exploration and FD only for chain-event
refinement. Details in~\refSup{sm::AbGrad}.

\begin{table}[h]
\centering
\caption{Gradient strategy ablation on vehicle chain-collision scene S6
($K{=}5$). Ev.\,1: car--trashbin; Ev.\,2: trashbin--table.}
\label{tab:ablation-grad}
\setlength{\tabcolsep}{5pt}\small
\renewcommand{\arraystretch}{0.80}
\begin{tabular}{lcccr}
\toprule
Method & Ev.\,1 & Ev.\,2 & Completion & Avg Time\,(s)\\
\midrule
AD-only (Tape) & 5/5 & 0/5 & 50.0\% & \phantom{0}35.2\\
FD-only        & 5/5 & 4/5 & 90.0\% & 153.5\\
\textbf{Hybrid (Ours)} & \textbf{5/5} & \textbf{4/5} & \textbf{90.0\%} & \textbf{101.3}\\
\bottomrule
\end{tabular}
\end{table}

\paragraph{Ablation Study of Collision Model.}
We compare simple-proxy optimization against proxy\,+\,SDF refinement on scene S6 ($K{=}10$), measuring event-2 (trashbin-table collision) success rate. The simple-proxy method achieves only 1/10 (10\%): coarse geometry cannot reliably capture the contact direction needed to guide the trashbin toward the table. Adding SDF refinement raises success to 8/10 (80\%), as accurate surface-distance gradients substantially improve chain-collision steering. Details are in~\refSup{sm::AbSDF}.
\section{Conclusion and Discussion}

We present a framework for generating executable, physically consistent dynamic 3D scenes from natural language in Unreal Engine. Central to our approach is the Evolutive Scene Graph (ESG), which encodes entity properties, spatial layout, and event timelines in a machine-checkable form. Gradient-based optimization grounds semantic descriptions into concrete layouts, and differentiable physics optimization refines physical parameters to satisfy the specified event timeline. Experiments confirm substantial improvements over baselines in event completion and parameter accuracy.

\paragraph{Limitations and Future Work.}
The system currently targets rigid-body dynamics and does not handle deformables, fluids, or articulated motion. Discrepancies between proxy geometry and UE runtime collision shapes also limit high-precision simulation. Future work will address richer physical parameterizations, tighter proxy--runtime alignment, and extensions to human actions and embodied agents.


\bibliography{aaai2027}

\clearpage
\appendix

\twocolumn[
\begin{center}
    {\huge \bf  Generating Physically Consistent Dynamic Scenes from Text Descriptions}\\
    {\LARGE \emph{Supplementary Material}}
\end{center}
]

\section{Technical Implementation Details}

\subsection{Layout Optimization Details}

\label{sm::layopt}

\paragraph{Pose parameterization and feasibility.}
We optimize a lightweight pose for each object $o_i$ as $\mathbf{x}_i=(\mathbf{p}_i,\mathbf{r}_i)$, where $\mathbf{p}_i\in\mathbb{R}^3$ is position and $\mathbf{r}_i\in\mathbb{R}^3$ is orientation.
Each object is assigned a semantic region $R(o_i)\in \{R_r\}_{r\in\mathcal{R}}$.
Hard feasibility (e.g., support height induced by \texttt{[On]}) is enforced by projection after each update, while the remaining constraints are modeled as differentiable penalties.
For SDF-based collision handling, we augment each object with optional geometry fields: a dense SDF grid and a proxy point set sampled near the object surface .

\paragraph{Relation energy and multi-relation aggregation.}
As defined in the main text, we represent semantic relations as a typed directed graph $\mathcal{G}=(\mathcal{O},\mathcal{E})$.
Each edge $e=(o_i\!\rightarrow\!o_j,\tau)\in\mathcal{E}$ contributes an unweighted edge penalty $E_{e}(\mathbf{x}_i,\mathbf{x}_j)=F_{\tau}(\mathbf{x}_i,\mathbf{x}_j)$.
When multiple relations exist between a pair of objects, we keep them as multiple edges and sum their contributions.
We aggregate relation energies over all incident edges in the layout to obtain the global relation energy:
\begin{equation}
E_{\mathrm{rel}}(\mathbf{O})=\sum_{e=(o_i\rightarrow o_j,\tau)\in\mathcal{E}} w_{e}F_{\tau}(\mathbf{x}_i,\mathbf{x}_j).
\end{equation}

\paragraph{Examples of Penalty Calculation.}
We denote by $\mathcal{B}_i(\mathbf{x}_i)\subset\mathbb{R}^3$ the collision proxy of object $o_i$ under pose $\mathbf{x}_i$, which can be instantiated by either a simple primitive (e.g., sphere/AABB/OBB) or a more faithful composed proxy when enabled (e.g., SDF). In our solver, each relation edge $e$ contributes to the total energy via a weighted penalty $w_e\,F_\tau$, where $w_e=\text{strictness}(\tau)\cdot\text{confidence}(e)$ and $F_\tau\ge 0$ is the unweighted violation defined by the corresponding constraint.

While simple proximity relations like \textit{Near} use a straightforward dead-zone penalty on surface distance $d_{\mathrm{surf}}(\mathcal{B}_i,\mathcal{B}_j)$ (as illustrated in the main text), directional relations require decomposing the geometry along specific axes.

(\emph{Near})
As an intuitive example, to instantiate $F_{\texttt{near}}$, we evaluate the surface distance $d_{\mathrm{surf}}$ of the collision proxies of the two objects. 
Given an adaptive preferred distance $d^\star$ and a tolerance $\delta$, we apply a dead-zone penalty:
\begin{equation}
F_{\texttt{near}}(\mathbf{x}_i,\mathbf{x}_j) = \left[d_{\mathrm{surf}} - (d^\star+\delta)\right]_+ + \left[(d^\star-\delta) - d_{\mathrm{surf}}\right]_+,
\end{equation}

In implementation, the adaptive preferred distance $d^\star$ is computed from both object scale($\bar{s}_{i/j}$)and semantic-region scale($S_{R}$):
\begin{equation}
d^\star=\operatorname{clamp}\!\left(
\min\!\left(
\frac{1}{2}\cdot \min(\bar{s}_i,\bar{s}_j),\;
\rho_d \, S_{R(i)}
\right),\;
d_{\min},\; d_{\max}
\right),
\end{equation}

where $[z]_+ = \max(z, 0)$, \(\bar{s}_k=\frac{4d_k^x+4d_k^y+d_k^z}{9}\) is the characteristic size of object \(o_k\),
\(S_{R(i)}=\sqrt{(u_{R(i)}^x-l_{R(i)}^x)^2+(u_{R(i)}^y-l_{R(i)}^y)^2}\) is the XY size of the source semantic region \(R(i)\),
and \(\rho_d\) is the distance ratio.

In our default setup, \(\rho_d=0.1\), \((d_{\min},d_{\max})=(8,300)\,\mathrm{cm}\), and
\(\delta=\operatorname{clamp}(\rho_t d^\star,\delta_{\min},\delta_{\max})\) with \(\rho_t=0.2\), \((\delta_{\min},\delta_{\max})=(2,80)\,\mathrm{cm}\).This yields zero penalty inside the admissible band $[d^\star-\delta, d^\star+\delta]$ and applies a linear penalty for outside. 

(\emph{Left\_of})
We define directional relations in the global coordinate system, where \textit{left} corresponds to $+X$.
Let $\Delta_x = x_j-x_i$ be the primary-axis offset (positive means the target is on the required side) and $\Delta_y = |y_j-y_i|$ be the lateral deviation on the orthogonal axis.
We decompose the violation into a primary term and a lateral term:
\begin{equation}
F_{\textit{left\_of}}(\mathbf{x}_i,\mathbf{x}_j)=F^{x}_{\textit{left\_of}}(\Delta_x)+F^{y}_{\textit{left\_of}}(\Delta_y).
\end{equation}
The primary-axis term is modeled as a piecewise penalty that strongly penalizes wrong-side configurations and softly penalizes offsets outside a reasonable range:
\begin{equation}
F^{x}_{\textit{left\_of}}(\Delta_x)=
\begin{cases}
\phi_{\mathrm{wrong}}(\Delta_x), & \Delta_x<0,\\
\phi_{\mathrm{near}}(\Delta_x), & 0\le \Delta_x<\delta_{\min},\\
0, & \delta_{\min}\le \Delta_x\le \delta_{\max},\\
\phi_{\mathrm{far}}(\Delta_x), & \Delta_x>\delta_{\max},
\end{cases}
\end{equation}
where \(\delta_{\min}\) is an adaptive minimum offset, \(\delta_{\max}\) is an adaptive upper range bound. Here \(\phi_{\mathrm{wrong}}, \phi_{\mathrm{near}}, \phi_{\mathrm{far}}\) denote three non-negative penalty functions (with different shapes/strengths) used to realize the primary-axis piecewise design: they map the primary offset \(\Delta_x\) to a violation magnitude for the three regimes (wrong side, insufficient offset, and excessive offset), while the preferred range \([\delta_{\min},\delta_{\max}]\) incurs zero penalty.

The lateral-alignment term applies a quadratic penalty beyond an adaptive tolerance \(t_y\):
\begin{equation}
F^{y}_{\textit{left\_of}}(\Delta_y)=\frac{1}{c}\left[\Delta_y-t_y\right]_+^2,
\end{equation}
with \(c>0\) a scaling constant. 
The same design extends to other directional constraints (e.g., \textit{In\_front\_of}) by pairing a primary-axis term with an orthogonal-axis alignment term.

\paragraph{Collision and region penalties.} 
\modi{This section has been revised comprehensively.}
Let \(\mathcal{P}_{\mathrm{all}}\) be all valid pairs \((i,j)\) with \(i<j\), and let \(L_i\in\{0,1\}\) indicate whether object \(i\) is treated as a large collision object (\(L_i = 1\) when mesh complexity above a threshold). We define the following collision energy :
\begin{equation}
\begin{aligned}
E_{\mathrm{col}}(\mathbf{O}) =\;
  &\sum_{(i,j)\in\mathcal{P}_{\mathrm{aabb}}}\psi_{\mathrm{aabb}}(i,j)
  +\sum_{(i,j)\in\mathcal{P}_{\mathrm{sdf}}}\psi_{\mathrm{sdf}}(i,j)\\
  &+\sum_{(i,j)\in\mathcal{P}_{\mathrm{hybrid}}}\psi_{\mathrm{hybrid}}(i,j).
\end{aligned}
\end{equation}

\textbf{AABB term.} For axis \(a\in\{x,y,z\}\), define overlap depth
\[
\delta_{ij}^{a}=
\left[\min(b_{i,\max}^{a},b_{j,\max}^{a})-\max(b_{i,\min}^{a},b_{j,\min}^{a})\right]_+.
\]
Then
\begin{equation}
\psi_{\mathrm{aabb}}(i,j)=
w_{\mathrm{aabb}}\sum_{a\in\{x,y,z\}}\left(\delta_{ij}^{a}\right)^2,
\end{equation}
where \(w_{\mathrm{aabb}}=\texttt{ANALYSIS\_AABB\_WEIGHT}\).

\textbf{SDF and HYBRID term.} Let \(\mathcal{Q}_{i\rightarrow j}\) be query samples on object \(i\) evaluated against object \(j\), and similarly \(\mathcal{Q}_{j\rightarrow i}\). For \(\mathbf{q}\in\mathcal{Q}_{i\rightarrow j}\), define directional penetration
\[
p_{i\rightarrow j}(\mathbf{q})=\left[m-d_j(\mathbf{q})\right]_+,
\]
where \(m\) is safety margin and \(d_j(\mathbf{q})\) is signed distance to object \(j\). 

For \(\mathcal{P}_{\mathrm{sdf}}\), both objects use SDF queries. For \(\mathcal{P}_{\mathrm{hybrid}}\), one object uses SDF queries and the large-object side uses signed distance to AABB. 

Define active (penetrating) sets :
\begin{align*}
\mathcal{H}_{i\rightarrow j} &= \{\mathbf{q}\in\mathcal{Q}_{i\rightarrow j}\mid p_{i\rightarrow j}(\mathbf{q})>0\},\\
\mathcal{H}_{j\rightarrow i} &= \{\mathbf{q}\in\mathcal{Q}_{j\rightarrow i}\mid p_{j\rightarrow i}(\mathbf{q})>0\}.
\end{align*}

and three disjoint pair subsets :
\begin{align*}
\mathcal{P}_{11} &=\{(i,j)\in\mathcal{P}_{\mathrm{all}}\mid L_i=L_j=1\}, \\
\mathcal{P}_{00} &=\{(i,j)\in\mathcal{P}_{\mathrm{all}}\mid L_i=L_j=0\}, \\
\mathcal{P}_{01} &=\{(i,j)\in\mathcal{P}_{\mathrm{all}}\mid L_i\neq L_j\}.
\end{align*}
The runtime pair partition is
\begin{equation}
\big(\mathcal{P}_{\mathrm{aabb}},\mathcal{P}_{\mathrm{sdf}},\mathcal{P}_{\mathrm{hybrid}}\big)=
\begin{cases}
(\mathcal{P}_{\mathrm{all}},\varnothing,\varnothing), & \texttt{AABB},\\
(\varnothing,\mathcal{P}_{\mathrm{all}},\varnothing), & \texttt{SDF},\\
(\mathcal{P}_{11},\mathcal{P}_{00},\mathcal{P}_{01}), & \texttt{HYBRID}.
\end{cases}
\end{equation}

Define
\[
S_{ij}=
\sum_{\mathbf{q}\in\mathcal{H}_{i\rightarrow j}} p_{i\rightarrow j}(\mathbf{q})^2+
\sum_{\mathbf{q}\in\mathcal{H}_{j\rightarrow i}} p_{j\rightarrow i}(\mathbf{q})^2,\quad \\
N_{ij}=|\mathcal{H}_{i\rightarrow j}|+|\mathcal{H}_{j\rightarrow i}|.
\]
So finally the common pair penalty is :
\begin{equation}
\psi_{\star}(i,j)=
\begin{cases}
w_{\mathrm{sdf}}\,\dfrac{S_{ij}}{N_{ij}}, & N_{ij}>0,\\[4pt]
0, & \text{otherwise},
\end{cases}
\end{equation}
where \(w_{\mathrm{sdf}}=\texttt{ANALYSIS\_SDF\_WEIGHT}\), with \(\psi_{\mathrm{sdf}}=\psi_{\star}\) on \(\mathcal{P}_{\mathrm{sdf}}\), \(\psi_{\mathrm{hybrid}}=\psi_{\star}\) on \(\mathcal{P}_{\mathrm{hybrid}}\) .

Region penalty is applied to object centers against semantic-region bounds:
\begin{equation}
E_{\mathrm{reg}}(\mathbf{O})=
w_{\mathrm{reg}}\sum_i\sum_{a\in\mathcal{A}_i}
\left(
[l_i^a-p_i^a]_+^2+[p_i^a-u_i^a]_+^2
\right),
\end{equation}
where \(\mathbf{p}_i=(p_i^x,p_i^y,p_i^z)\) is the center position of object \(i\), and \(p_i^a\) denotes its coordinate on axis \(a\); while \([l_i^a,u_i^a]\) is the valid interval on axis \(a\).

\paragraph{Preference strategy in initial placement.}
Beyond constraint satisfaction, we incorporate a lightweight \emph{preference energy} to improve the aesthetic quality and practicality of the initial draft. For each object, the LLM predicts a coarse preferred direction within its assigned semantic region (e.g., center, near boundary, left/right/front/back). We discretize the region into a sampling grid and construct a preference cost map \(E^{\text{pref}}_i(\mathbf{p})\) over candidate positions \(\mathbf{p}\) by measuring the deviation from the preferred direction (or anchor) using a simple spatial prior. The overall score used in grid sampling is
\[
E_i(\mathbf{p}) \;=\; E^{\text{cons}}_i(\mathbf{p}) \;+\; \lambda_{\text{pref}}\, E^{\text{pref}}_i(\mathbf{p}),
\]
where \(E^{\text{cons}}_i\) aggregates relational and feasibility terms (e.g., collisions and region validity), and \(\lambda_{\text{pref}}\) keeps preferences secondary to hard/strong constraints. This design makes the placer prioritize constraint satisfaction when constraints are present, while using preferences to break symmetries and select semantically reasonable positions when constraints are weak or underspecified.

In practice, adding \(E^{\text{pref}}\) yields more balanced drafts (e.g., plates evenly distributed around a table, with a vase and cup holder placed near the center), whereas removing it often produces visually unbalanced layouts because the sampler has no cue to distribute unconstrained objects. A better initial draft also tends to reduce early collisions and provides a more favorable starting point for subsequent continuous optimization, improving both convergence stability and speed.
This preference strategy produces a low-conflict draft quickly; detailed anti-penetration is delegated to the later continuous stage. Keeping preference lightweight avoids introducing unnecessary geometric bias before relation constraints are stabilized.

\paragraph{Optimization and projection.}
As formulated in the main text, the total energy is given by:
\begin{equation}
E(\mathbf{O})=E_{\mathrm{rel}}(\mathbf{O})+\lambda_{\mathrm{col}}E_{\mathrm{col}}(\mathbf{O})+\lambda_{\mathrm{reg}}E_{\mathrm{reg}}(\mathbf{O})
\end{equation}
where $E_{\mathrm{rel}}(\mathbf{O})$ aggregates relation-edge energies on the global graph as discussed above. The collision term $E_{\mathrm{col}}(\mathbf{O})$ and the region-boundary term $E_{\mathrm{reg}}(\mathbf{O})$ are defined in Sec.~\emph{Collision and region penalties}.

We minimize the total energy with Adam updates and gradient clipping, followed by projection $\mathrm{Proj}_{\mathcal{D}}(\cdot)$ to enforce region bounds and hard constraints:
\begin{equation}
\boldsymbol{\theta}^{(t+1)}=
\mathrm{Proj}_{\mathcal{D}}\!\left(
\boldsymbol{\theta}^{(t)}-\alpha\frac{\hat{\mathbf{m}}^{(t)}}{\sqrt{\hat{\mathbf{s}}^{(t)}}+\epsilon}
\right).
\end{equation}
We stop when the relative energy decrease falls below $\epsilon_{\mathrm{stop}}$ or a maximum number of iterations is reached.

\subsection{Physical Optimization Details}
\label{sm::phyopdt}

\paragraph{Optimizable variables and bounds.}
We collect all learnable timeline variables into a parameter vector $\boldsymbol{\theta}$. Each entry corresponds to a named attribute in the ESG---such as an actor's initial speed, a contact impulse magnitude, a damping coefficient, or an object mass---and is associated with a valid range $[\theta_{\min}, \theta_{\max}]$ inferred from the LLM-generated bounds in the timeline.
Parameters marked as fixed scalars in the ESG are excluded via a binary mask; after each optimizer step we project by clamping $\boldsymbol{\theta}$ back to its valid range to ensure physical feasibility throughout optimization.

\paragraph{Window loss.}
\label{sm::winloss}
The temporal window constraint $\mathcal{L}^e_\text{win}$ in Eq.~(11) is instantiated for the best-approach frame $\hat{t}$ (Eq. (13)) as a scale-normalized smooth barrier:
\begin{equation}
  \mathcal{L}_\text{win} =
    \left(\frac{\operatorname{softplus}\!\bigl(\beta(t_\text{lo}-\hat{t})\bigr)}
               {\beta\,\bigl(|t_\text{hi}-t_\text{lo}|+\varepsilon\bigr)}\right)^{\!2}
  + \left(\frac{\operatorname{softplus}\!\bigl(\beta(\hat{t}-t_\text{hi})\bigr)}
               {\beta\,\bigl(|t_\text{hi}-t_\text{lo}|+\varepsilon\bigr)}\right)^{\!2},
\end{equation}
where $\beta$ controls boundary sharpness and $\varepsilon$ is a numerical stabilizer. For events with a fixed target time (i.e., $t_\text{lo} = t_\text{hi}$), the denominator collapses to $\varepsilon$ and the term degrades to a squared deviation from the target moment.

\paragraph{Loss terms for additional event types.}
\label{sm::eventtypes}
Beyond the \textsc{Collision} type formulated in Eq.~(12--13), we define the following losses for the remaining event types.

\textit{PositionAt.}
This point event requires that object $i$ occupies a specified position
$\mathbf{p}^*$ at time $t_p$:
\begin{equation}
  \mathcal{L}^\text{POS}_\text{task}
    = \bigl\|\mathbf{p}_i(t_p) - \mathbf{p}^*\bigr\|_2^2.
\end{equation}

\textit{UniformMotion and Brake (interval events).}
An interval event $i = (u,v) \in \mathcal{I}$ prescribes a motion regime over the span between bounding point events $u$ and $v$.
For a \textsc{UniformMotion} segment with target speed $v^*$, we penalize deviation from constant-speed travel at sampled interior frames
$\{t_k\}$:
\begin{equation}
  \mathcal{L}^\text{UNI}_\text{task}
    = \frac{1}{K}\sum_{k=1}^{K}
        \bigl(\|\dot{\mathbf{p}}_i(t_k)\| - v^*\bigr)^2.
\end{equation}
For a \textsc{Brake} segment we modify the linear damping of the object, in order to simulate the braking motion.
All interval-event losses use the same window-loss structure as point
events, with $[t_\text{lo}, t_\text{hi}]$ set to the span of the
enclosing interval.

\paragraph{Gradient computation and two-phase strategy.}
The AD surrogate integrates a penalty-based soft-contact impulse
$\mathbf{f} = k_e\max(0, -d)\hat{\mathbf{n}}$ at each timestep,
where $d$ is the signed surface distance between proxy shapes and
$\hat{\mathbf{n}}$ the contact normal.
For sphere proxies the contact normal is the exact center-to-center
direction, making the surrogate gradient reliable for direct
single-body interactions.
For non-spherical proxies (boxes, capsules) the normal is approximated
from the primitive SDF, which introduces directional error that
accumulates across chain interactions; we therefore switch to FD
gradients on the full Newton simulator whenever the optimizer detects
a chain event (i.e., the active body for event $e_{k+1}$ is a
\emph{passive} body in event $e_k$).
The switch threshold is determined statically from the ESG dependency
graph at planning time, requiring no runtime detection.

\paragraph{Implementation and parameter update.}
We implement the forward rollout and loss evaluation in NVIDIA Warp and
differentiate via \texttt{wp.Tape} (reverse-mode AD).
Gradient updates follow Adam with step size $\eta$:
\begin{equation}
  \boldsymbol{\theta}^{(k+1)}
    = \Pi_{[\boldsymbol{\theta}_{\min},\boldsymbol{\theta}_{\max}]}
      \!\Bigl(\boldsymbol{\theta}^{(k)}
              - \eta\,\widetilde{\nabla}_{\!\boldsymbol{\theta}} L\Bigr),
\end{equation}
where $\widetilde{\nabla}$ denotes the stabilized gradient after masking
fixed entries, sanitizing NaN/Inf values, and clipping by global norm,
and $\Pi$ projects parameters back into their valid range.
A coarse grid search over a $20$-point discretization of $\boldsymbol{\theta}$
precedes Adam to provide a warm-start initialization, reducing
sensitivity to the initial parameter draw.

\subsection{Scene Class Generation Details}
\label{sm::clsgen}

Given a validated ESG draft and the resolved asset bindings, our backend deterministically translates the structured specification into an engine-executable scene class that can be directly integrated into an Unreal Engine project. 

\paragraph{Atomic control primitives.}
High-level motion intents produced by the optimization layer (e.g., \texttt{[Uniform Motion]} or \texttt{[Braking]}) are not directly executable in UE without an explicit control interface. We therefore represent dynamics as a sequence of atomic control primitives that map one-to-one onto UE operations. Concretely, we define primitives such as \texttt{[SetVelocity]}, \texttt{[AdjustProperties]}, and \texttt{[ApplyImpulse]}, and implement a deterministic translation from abstract motions to these primitives. This design provides a strict mapping from event-level intent to low-level, physics-executable control, making dynamic scenes reproducible, debuggable, and easy to iterate in a real engine.

\paragraph{Scene instantiation and code emission.}
The generated class is responsible for loading assets, instantiating and placing objects, initializing physics states, and executing dynamic events according to the scene timeline. During instantiation, we set key UE runtime properties such as \texttt{[Mobility]}, \texttt{[SimulatePhysics]}, \texttt{[Mass]}, and \texttt{[Damping]} to ensure the scene behaves consistently under physics simulation.

For static layout, we instantiate each object according to its ESG-specified pose and scale, and load the corresponding \texttt{[StaticMesh]} and \texttt{[Materials]} from the resolved asset paths. For dynamics, we consume the ESG timeline as an ordered list of atomic events; each event records its trigger time, target objects, primitive type, and parameters.

At runtime, the generated class maintains a simulation clock and, within UE's Tick loop, checks whether the current time reaches the trigger time of any pending events. When an event is due, a dispatcher invokes the corresponding UE API calls to apply the specified atomic controls. For transparency and debugging, the system emits structured messages to UE's Output Log at the onset of each high-level event, allowing users to track event execution and inspect relevant runtime states. In addition, scene parameters remain editable in the UE editor, enabling users to interactively adjust configurations and iterate on the scene efficiently.

\section{Additional Evaluation and Details}

\subsection{Time Consumption Evaluation}
\label{sm::time}

On a single workstation, generating one scene takes about 350 seconds end to end on average. The runtime is primarily distributed across three stages, LLM drafting, which accounts for approximately 120 seconds. Asset resolution, which takes around 30 seconds and covers both retrieval and optional generation. Layout and physical optimization, which requires about 200 seconds, related to the complexity of the scene layout and timeline arrangement.
\subsection{More Quality Assessment}
\label{sm::assessment}
\paragraph{Specifying asset properties.}

\begin{figure*}[t]
    \centering
    \includegraphics[width=\linewidth]{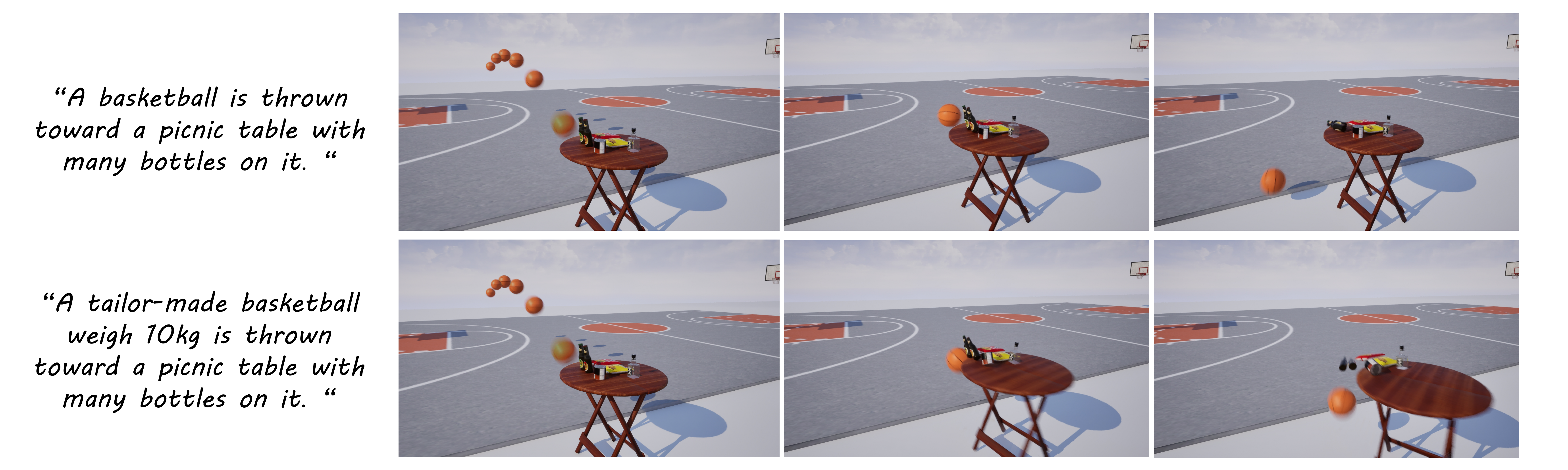}
    \caption{Scene generation with different asset properties description.
    We vary the mass of the basketball, and set other factors the same,
    leading to evidently different situations.}
    \label{experiment::diff_mass}
\end{figure*}

~\refFig{experiment::diff_mass} demonstrates the system's ability to respect explicit physical property constraints embedded in the prompt. Both rows depict a basketball thrown toward a picnic table with bottles; the only difference is the mass specification: the first prompt uses a default basketball, while the second specifies a \emph{tailor-made basketball weighing 10\,kg}. With the heavier mass, the optimizer assigns a correspondingly higher momentum to satisfy the same collision event, resulting in a noticeably more forceful impact---the table and bottles are displaced more dramatically, as visible in the final frames. All other scene factors (initial position, target, trajectory) remain identical, isolating mass as the sole variable. This confirms that the system correctly propagates LLM-inferred physical priors into the optimization loop, producing outcomes that are both semantically faithful and physically differentiated by the specified asset properties.

\begin{figure*}[b]
    \centering
    \includegraphics[width=\linewidth]{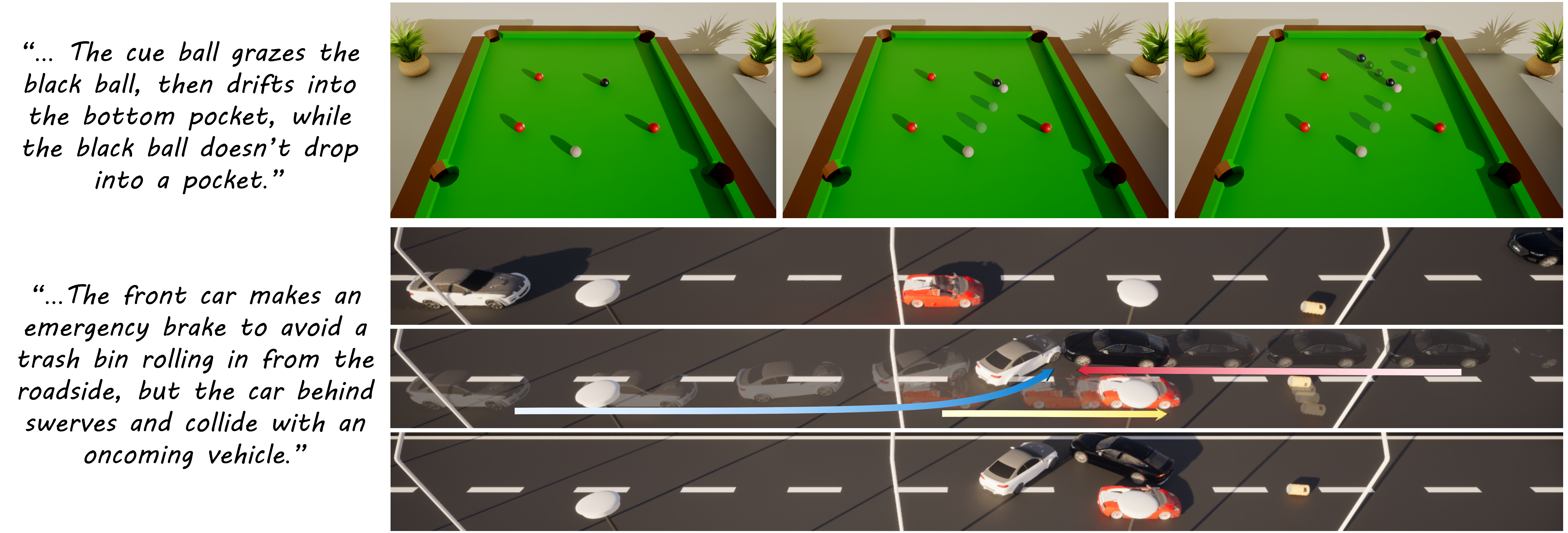}
    \caption{Scene generation with different event description.
    The first case is an adaptation of the first case
    in~\refFig{experiment::cascade}, and the second case is an
    adaptation of the second case in~\refFig{experiment::long_sequence}.}
    \label{experiment::diff_event}
\end{figure*}

\paragraph{Assessing prompt sensitivity.}
We further probe prompt sensitivity through two controlled experiments. First, keeping the scene unchanged and modifying only object properties (~\refFig{experiment::diff_mass}): the altered specification yields noticeably different motion, confirming that the system respects asset-level physical constraints. 
Second, fixing assets and varying only the event description (~\refFig{experiment::diff_event}): changing which ball to pocket produces qualitatively distinct, intent-aligned outcomes, confirming sensitivity to subtle motion-semantics edits. 

\subsection{Comparison Experiment Details}
\label{sm::comparison}

\paragraph{Test scenes.}
\refTab{tab:scenes} lists all 10 test prompts together with their
expected events and explicit quantitative constraints.
Scenes S1--S4 each impose a single contact event;
S5--S7 require two chained contacts with multi-step momentum transfer;
S8--S10 involve more than one behavioral phase with concurrent or
time-windowed constraints.
Five scenes (S3, S4, S7, S8, S10) carry explicit quantitative
targets---event timing, output velocity, or object mass---used for
the parameter accuracy evaluation in Table~\ref{tab:params}.

\paragraph{Evaluation protocol.}
\label{sm::metrics}
We evaluate three complementary metrics derived from the engine-native
simulation log.

\textbf{Scene Layout Rate (SLR)} measures static initialization
validity before simulation begins.
A scene is valid when (i)~all text-prescribed objects are successfully
spawned, and (ii)~no pair of objects exhibits an initial overlap depth
exceeding $\delta{=}2$\,cm, checked via UE's
\texttt{ComputePenetration} API\@.
Each of the $K{=}5$ runs uses an independent random seed; the small
but non-zero variance across runs reflects stochastic elements in
LLM-driven asset placement.
Inspired by the Scene Validity Rate of FATE~\cite{li2025fate},
SLR separates layout quality from dynamic success.

\textbf{Event Completion Rate (ECR)} is an engine-native binary
criterion for each prescribed dynamic event,
analogous to the Execution Validity Rate of FATE and the Task
Success Rate of WorldArena~\cite{worldarena2025}.
An event is counted as completed only when all four conditions are
simultaneously satisfied:
(1)~the target \texttt{OnComponentHit} callback fires within
$T_{\max}{=}15$\,s;
(2)~penetration depth at the moment of contact is less than 5\,cm
(ruling out initial-overlap false triggers);
(3)~normal impulse exceeds $F_{\min}{=}0.1$\,N$\cdot$s
(ruling out ghost contacts);
(4)~a re-sample 0.2\,s post-hit confirms that the objects are no
longer penetrating beyond 2\,cm (ruling out continued tunneling).
For chain events, later-stage contacts are credited only when all
preceding contacts have already been completed in the correct order,
with a minimum inter-event gap of 0.1\,s to prevent same-frame
double-counting.

\textbf{Physical Parameter Accuracy (PPA).}
\label{sm::ppa_definition}
PPA measures how closely prompt-specified parameters are satisfied,
using a continuous score in $[0,1]$ per parameter instance rather than
a binary pass/fail.

\textit{Score definition.}
For a \textbf{point constraint} $x{=}x^*$ with precision threshold
$\epsilon$, the per-instance score is
\[
  s(x,\,x^*) \;=\;
  \begin{cases}
    1 & |x - x^*| \le \epsilon,\\[4pt]
    \displaystyle\max\!\left(0,\;
      1 - \frac{|x - x^*| - \epsilon}{x^*}\right)
      & \text{otherwise.}
  \end{cases}
\]
The denominator $x^*$ normalises the violation by the magnitude of
the target, so that a given absolute error is penalised more for
smaller targets.
For an \textbf{interval constraint} $x\in[a,b]$, let
$e(x)=\max(0,\,a-x,\,x-b)$ be the out-of-interval excess; the score is
\[
  s(x,\,[a,b]) \;=\;
  \begin{cases}
    1 & e(x)=0,\\[4pt]
    \displaystyle\max\!\left(0,\;
      1 - \frac{2\,e(x)}{b-a}\right)
      & \text{otherwise.}
  \end{cases}
\]
Here the half-width $(b-a)/2$ serves as the reference scale: an
excess equal to the half-width yields a score of zero.
In both cases a score of~$1$ indicates full satisfaction;
a score in $(0,1)$ indicates a proportional violation; a score of~$0$
indicates a violation at least as large as the reference scale.

\textit{Aggregation.}
For initial-condition parameters (mass, initial velocity) directly
set at scene initialisation, all $K{=}5$ runs contribute to the
average regardless of event completion.
For outcome parameters (collision timing, post-collision velocity)
that require a dynamic event to realise, runs where the target event
does not complete receive a score of~$0$.
PPA is the mean score over all parameter instances and runs.
We report PPA under three precision thresholds in
Table~\ref{tab:ppa_sensitivity} and show that all qualitative
conclusions are robust.
Scene Language is excluded as it outputs no physical parameters.
The underlying absolute errors are reported in
Table~\ref{tab:params}.

\textbf{Ablation baseline (w/o Opt.)} is a variant of our system
in which the differentiable physical optimization stage is entirely
removed.
Physical parameters (initial velocities, forces, and masses) are
retained at their LLM-assigned heuristic values and remain fixed
throughout simulation; no gradient-based refinement is performed.
The static layout optimization, asset retrieval, ESG compilation, and
UE instrumentation pipeline are identical to our full method, so SLR
is unaffected.
This variant isolates the contribution of differentiable event-loss
optimization to ECR and PPA.

Each method is run $K{=}5$ times per scene with independent random
seeds.
Static methods (SceneCraft, PAT3D) are not UE-executable and are
excluded from the formal comparison; they are discussed in the main
text as representative layout baselines.
Scene Language is evaluated by importing extracted primitive data
into UE via our three-stage conversion pipeline
(Stage~1: shape extraction; Stage~2: velocity inference; Stage~3:
C++ class generation), and running full physics simulation with the
same ECR instrumentation applied to all other methods.
SimWorld is instrumented with the identical ECR callback logic.

\paragraph{Scene Language $\to$ UE adapter.}
\label{supp:sl-adapter}
Scene Language~\cite{zhang2025scenelanguage} outputs a program-based
representation without UE-native physics state.
We import it via a two-step adapter: (i)~execute the program in
\texttt{exposed} mode, extract primitive poses, group them into
semantic rigid bodies, and map them to UE BasicShapes under a fixed
coordinate transform
($(X,Y,Z)_{\mathrm{UE}}{=}100(-z,x,y)_{\mathrm{SL}}$~cm);
(ii)~infer a linear launch velocity from the last two kinematic-trail
centers with a scene-type reference speed, setting angular velocity to
zero.
The resulting JSON is compiled by the same C++ emitter and ECR
callbacks as our method.
Approximations (primitive proxies, fixed reference speed, no angular
velocity) can only understate Scene Language's visual quality; observed
failures are dominated by layout overlap (SLR\,$\approx$\,$72\%$)
rather than inflated event completion.

\paragraph{Quantitative results.}
Table~\ref{tab:completion} reports per-scene event completion for
Scene Language, SimWorld, \textbf{w/o Opt.}, and our full method
across the 10 test scenes.

\textit{Scene Layout Rate.}
Our method (and its ablated variant \textbf{w/o Opt.}) achieves
SLR\,$=$\,$94\%_{\pm5\%}$ across all 10 scenes; the layout
optimization pipeline is shared, so both variants produce identical
static initialization quality.
SimWorld yields SLR\,$=$\,$89\%_{\pm4\%}$; its city-scale asset
placement heuristics do not enforce precise object-level overlap
constraints, causing occasional violations.
Scene Language achieves SLR\,$=$\,$72\%_{\pm8\%}$; S6
consistently fails SLR due to directly measured initial overlaps of
$-24$\,cm and $-38$\,cm between car, trash can, and bench, and
several other scenes exhibit occasional penetration failures due to
stochastic LLM sampling, resulting in the higher variance.

\textit{Event Completion Rate.}
Scene Language achieves 3.2/18 overall ($\pm 3.3$).
On Level-1 single-event scenes, its procedural trail primitives
occasionally produce plausible ball trajectories, yielding partial
success on ball-type scenes
(S1, S2, S4: $0.4_{\pm0.5}$ each);
car-type and timing-constrained scenes fail entirely
(S3: $0.0_{\pm0.0}$).
On chain events, Scene Language partially succeeds on S7
($0.8_{\pm0.7}$), where the 2\,cm initial ball spacing enables
event$_1$ completion via the inferred trail velocity; S6 scores 0
as its initial overlap prevents valid collision impulses.
Level-3 scenes all fall at or below $0.6/N$: the representation
encodes no multi-phase temporal structure, and post-event
snapshot initialization degrades layout validity further.

SimWorld completes 1.4/18, contributing only S6 ($0.4_{\pm0.5}$)
and S8 ($1.0_{\pm0.0}$), where pre-defined city-agent animations
coincidentally satisfy the event constraints; all other scenes score 0.

\textbf{w/o Opt.} achieves 8.8/18\,($\pm$2.1), revealing a clear
and interpretable failure pattern.
On Level-1 scenes, LLM-assigned heuristic parameters reliably
realize all simple collision events where the trajectory is
determined by initial position and velocity alone
(S1, S2, S3: $1.0_{\pm0.0}$); S4 partially succeeds
($0.6_{\pm0.5}$) because the non-standard mass ($10$\,kg) combined
with a tight 1-second timing constraint increases initialization
sensitivity and reduces event realization rate.
On chain events (S5--S7), the first contact succeeds in most runs
(S5: $0.8_{\pm0.4}$; S6: $0.6_{\pm0.5}$; S7: $1.0_{\pm0.0}$),
but the second contact fails in \emph{all} runs without exception:
no gradient feedback steers the post-collision trajectory of the
intermediate body toward its downstream target.
On Level-3 multi-phase scenes, the first event succeeds when it
involves straightforward rectilinear motion (S8, S10 event~1), but
turning maneuvers fail in most runs (S9 event~1: $0.4_{\pm0.5}$) and
braking sequences are correctly initiated in at most two runs (S8
event~2: 2/5; S10 event~2: 1/5), leaving all downstream collisions
unrealized.
The 7.6-point ECR gap relative to our full method (8.8 vs.\ 16.4)
directly quantifies the contribution of differentiable event-loss
optimization.

Our full method achieves 16.4/18 ($\pm 1.0$).
The two incomplete events occur in S10, the most complex three-body
chain scene, where gradient signal attenuates over the extended
optimization horizon.
Pairwise Wilcoxon signed-rank tests (per-run sums, $K{=}5$) confirm
our method significantly outperforms all baselines ($p{<}0.05$).

\textit{Physical Parameter Accuracy.}
Table~\ref{tab:params} reports the underlying absolute errors;
PPA scores are computed from these errors per the definition
in Supp.~Mat.~\ref{sm::ppa_definition}.
Scene Language is excluded as it produces no physical parameters.

\textbf{w/o Opt.}\ achieves PPA\,$=$\,61.0\%\,($\pm$5\%),
revealing a qualitative split between parameter types.
\emph{Literal initial-condition parameters} directly assignable by
the LLM (mass, initial velocity, speed range) score~1.0 in all
runs: S4 mass error is $0.20_{\pm0.10}$\,kg, S8 initial speed
error is $0.40_{\pm0.30}$\,m/s, and S10 speed lies within
$[5,10]$\,m/s in every run.
\emph{Emergent outcome parameters} depending on unfolding dynamics
receive only partial credit: collision timing (S3:
$1.50_{\pm0.70}$\,s, score\,$=$\,0.60; S4: $0.85_{\pm0.40}$\,s
with two non-completions, score\,$=$\,0.27) and post-collision
velocity (S7 $v_B$: $1.80_{\pm0.60}$\,m/s, score\,$=$\,0.60)
incur substantial deductions, while S10 timing fails entirely
(no completions, score\,$=$\,0).
Aggregated over outcome parameters only, \textbf{w/o Opt.}\
achieves an outcome PPA of $37\%$.

SimWorld achieves PPA\,$=$\,15.0\%\,($\pm$8\%): partial credit
on S8 initial speed (error\,$=$\,$2.92$\,m/s, score\,$=$\,0.81)
and S8 braking time (excess\,$=$\,0.32\,s, score\,$=$\,0.36)
yield small non-zero contributions; all other parameters score~0
due to event non-completion.

Our full method achieves PPA\,$=$\,95.0\%\,($\pm$3\%), with
near-perfect scores on all constraints in all runs.
The three sub-unity contributions are the S8 velocity outlier
(error\,$=$\,1.1\,m/s in one run) and two S10 rear-end collision
non-completions.
The 34-point gap in overall PPA between \textbf{w/o Opt.}\
(61\%) and our method (92\%), and the 55-point gap on outcome
parameters alone (37\% vs.\ 92\%), confirm that differentiable
event-loss optimization is the decisive factor for satisfying
temporal and causal constraints.

\paragraph{PPA Precision Robustness.}
\label{sm::ppa_sensitivity}
The continuous PPA score uses a precision threshold $\epsilon$
to define the ``fully satisfied'' region (score\,$=$\,$1$); violations
beyond $\epsilon$ receive proportional partial credit.
Table~\ref{tab:ppa_sensitivity} reports PPA under three
threshold choices, confirming that all qualitative conclusions are
robust: (i)~the substantial gap between our method and \textbf{w/o
Opt.}\ on outcome parameters persists across all settings, and
(ii)~literal initial-condition parameters yield a stable score
floor for \textbf{w/o Opt.}\ that is largely threshold-independent.

\begin{table}[h]
\centering
\caption{PPA sensitivity to precision thresholds
(same 40 constraint measurements; Scene Language excluded).}
\label{tab:ppa_sensitivity}
\setlength{\tabcolsep}{4pt}
\small
\resizebox{\columnwidth}{!}{%
\begin{tabular}{lc ccc}
\toprule
Threshold set
  & $(\epsilon_t,\,\epsilon_v,\,\epsilon_m)$
  & w/o Opt. & SimWorld & \textbf{Ours} \\
\midrule
Strict          & $(0.2\,\text{s},\ 0.5\,\text{m/s},\ 0.3\,\text{kg})$
                & 57\%  & 14\%  & \textbf{95\%} \\
Standard (paper)& $(0.3\,\text{s},\ 1.0\,\text{m/s},\ 0.5\,\text{kg})$
                & 61\%  & 15\%  & \textbf{95\%} \\
Lenient         & $(0.5\,\text{s},\ 2.0\,\text{m/s},\ 1.0\,\text{kg})$
                & 68\%  & 16\%  & \textbf{95\%} \\
\bottomrule
\end{tabular}%
}
\end{table}

\begin{table}[!htb]
\centering
\caption{Test scene specifications. Events are listed in sequential
order; constraints in parentheses are the quantitative requirements
used in~\refTab{tab:params}.}
\label{tab:scenes}
\setlength{\tabcolsep}{4pt}
\small
\begin{tabular}{clp{6.8cm}c}
\toprule
ID & Level & Prompt \& Events & $N$ \\
\midrule
S1 & \multirow{4}{*}{\rotatebox[origin=c]{90}{\textit{Single}}}
   & A bowling ball rolls toward a set of bowling pins and hits them.
     \newline \textit{(1) ball--pin collision}
   & 1 \\
S2 & & A soccer ball hits a tidy office desk.
     \newline \textit{(1) ball--desk collision}
   & 1 \\
S3 & & A car move towards a jeep and collide with it after 3 seconds.
     \newline \textit{(1) car--jeep collision} $(t_{collide}{=}3.0\,\mathrm{s})$
   & 1 \\
S4 & & A tailor-made basketball weigh 10kg is thrown toward a picnic table with many bottles on it, and hit after 1 seconds.
     \newline \textit{(1) ball--table collision} $(m_{ball}{=}10\,\mathrm{kg},\; t_{collide}{=}1.0\,\mathrm{s})$
   & 1 \\
\midrule
S5 & \multirow{3}{*}{\rotatebox[origin=c]{90}{\textit{Chain}}}
   & There are several billiards on the billiard table. The cue ball strikes the black ball, sending it cleanly into the bottom pocket.
     \newline \textit{(1) cue--black collision \quad (2) black ball into pocket}
   & 2 \\
S6 & & A car collide with a trashbin, knocking the trashbin to collide with a picnic table by the street side.
     \newline \textit{(1) car--trashbin collision \quad (2) trashbin--table collision}
   & 2 \\
S7 & & There are three balls on the ground. Ball A moves and hits ball B, giving ball B an initial speed at 2m/s, then ball B hits ball C.
     \newline \textit{(1) A--B collision \quad (2) B--C collision} $(v_B{=}2.0\,\mathrm{m/s})$
   & 2 \\
\midrule
S8  & \multirow{3}{*}{\rotatebox[origin=c]{90}{\textit{Multi-phase}}}
    & A black car is moving at 10 m/s down the road. The driver starts braking, letting the car to stop in 2~3 seconds.
      \newline \textit{(1) car move down the road \quad (2) car brakes to stop}
      $(v_{car}{=}10\,\mathrm{m/s},\; t_\text{brake\_to\_stop}{\in}[2,3]\,\mathrm{s})$
    & 2 \\
S9  & & A car moves along the street and turns left at a cross. It brakes hard after entering the new lane, but still hits a bicycle stopped in the middle of the road.
      \newline \textit{(1) car turns left \quad (2) car brakes \quad (3) car--bicycle collision}
    & 3 \\
S10 & & Two cars moves at the same speed of 5~10m/s, one in front of the other. The front car suddenly brakes, and two cars collide after 1 second.
      \newline \textit{(1) two cars in motion \quad (2) front car brakes \quad (3) rear-end collision}
      $(v_{cars}{\in}[5,10]\,\mathrm{m/s},\; t_\text{collide}{=}1.0\,\mathrm{s})$
    & 3 \\
\bottomrule
\end{tabular}
\end{table}

\begin{table}[!htb]
\centering
\caption{Dynamic event completion on 10 test scenes, averaged over
$K{=}5$ independent runs. $N$ = expected events per scene.
Each cell shows mean\,$\pm$\,std completed events;
the Total row reports std over the five per-run sums.
\textbf{w/o Opt.} denotes our method without differentiable
physical optimization (LLM-assigned parameters only).}
\label{tab:completion}
\setlength{\tabcolsep}{4pt}
\small
\begin{tabular}{lc cccc}
\toprule
Scene & $N$ & Sc. Language & SimWorld & w/o Opt. & \textbf{Ours} \\
\midrule
\multicolumn{6}{l}{\textit{Level 1: Single event}} \\
S1  & 1 & $0.4_{\pm0.5}$ & $0.0_{\pm0.0}$ & $1.0_{\pm0.0}$ & $\mathbf{1.0_{\pm0.0}}$ \\
S2  & 1 & $0.4_{\pm0.5}$ & $0.0_{\pm0.0}$ & $1.0_{\pm0.0}$ & $\mathbf{1.0_{\pm0.0}}$ \\
S3  & 1 & $0.0_{\pm0.0}$ & $0.0_{\pm0.0}$ & $1.0_{\pm0.0}$ & $\mathbf{1.0_{\pm0.0}}$ \\
S4  & 1 & $0.4_{\pm0.5}$ & $0.0_{\pm0.0}$ & $0.6_{\pm0.5}$ & $\mathbf{1.0_{\pm0.0}}$ \\
\midrule
\multicolumn{6}{l}{\textit{Level 2: Chain events}} \\
S5  & 2 & $0.4_{\pm0.5}$ & $0.0_{\pm0.0}$ & $0.8_{\pm0.4}$ & $\mathbf{2.0_{\pm0.0}}$ \\
S6  & 2 & $0.0_{\pm0.0}$ & $0.4_{\pm0.5}$ & $0.6_{\pm0.5}$ & $\mathbf{1.8_{\pm0.4}}$ \\
S7  & 2 & $0.8_{\pm0.7}$ & $0.0_{\pm0.0}$ & $1.0_{\pm0.0}$ & $\mathbf{2.0_{\pm0.0}}$ \\
\midrule
\multicolumn{6}{l}{\textit{Level 3: Multi-phase actor dynamics}} \\
S8  & 2 & $0.0_{\pm0.0}$ & $1.0_{\pm0.0}$ & $1.2_{\pm0.4}$ & $\mathbf{2.0_{\pm0.0}}$ \\
S9  & 3 & $0.2_{\pm0.4}$ & $0.0_{\pm0.0}$ & $0.4_{\pm0.5}$ & $\mathbf{2.2_{\pm0.8}}$ \\
S10 & 3 & $0.6_{\pm0.5}$ & $0.0_{\pm0.0}$ & $1.2_{\pm0.4}$ & $\mathbf{2.4_{\pm0.8}}$ \\
\midrule
\textbf{Total} & \textbf{18}
  & $3.2_{\pm3.3}$
  & $1.4_{\pm0.5}$
  & $8.8_{\pm2.1}$
  & $\mathbf{16.4_{\pm1.0}}$ \\
\bottomrule
\end{tabular}
\end{table}

\begin{table}[!htb]
\centering
\caption{Absolute parameter errors underlying PPA,
on five scenes with explicit physical constraints.
Cell values are $|\Delta|_{\pm\text{std}}$;
$n/5$ denotes runs where the constraint is evaluated
(error is 0 when the output falls within a window constraint;
``\,---\,'' = event did not occur).
Initial-condition parameters are evaluated over all $K{=}5$ runs;
outcome parameters only over runs where the event completes,
with non-completion scored as~0.
\textbf{PPA (\%):}
w/o~Opt.\ 61.0;\; SimWorld 15.0;\; Ours 95.0.
Scene Language excluded (no physical parameters output).}
\label{tab:params}
\setlength{\tabcolsep}{3.0pt}
\small
\resizebox{\columnwidth}{!}{%
\begin{tabular}{cl ccc}
\toprule
 & Constraint & w/o Opt. & SimWorld & \textbf{Ours} \\
\midrule
S3
  & $t\;(\mathrm{s}) = 3.0$
    & $1.50_{\pm0.70}$ {\scriptsize(5/5)}
    & ---
    & $\mathbf{0.16_{\pm0.05}}$ {\scriptsize(5/5)} \\
\addlinespace
\multirow{2}{*}{S4}
  & $m\;(\mathrm{kg}) = 10$
    & $0.20_{\pm0.10}$ {\scriptsize(5/5)}
    & ---
    & $\mathbf{0.02_{\pm0.04}}$ {\scriptsize(5/5)} \\
  & $t\;(\mathrm{s}) = 1.0$
    & $0.85_{\pm0.40}$ {\scriptsize(3/5)}
    & ---
    & $\mathbf{0.06_{\pm0.05}}$ {\scriptsize(5/5)} \\
\addlinespace
S7
  & $v_B\;(\mathrm{m/s}) = 2.0$
    & $1.80_{\pm0.60}$ {\scriptsize(5/5)}
    & ---
    & $\mathbf{0.14_{\pm0.08}}$ {\scriptsize(5/5)} \\
\addlinespace
\multirow{2}{*}{S8}
  & $v\;(\mathrm{m/s}) = 10.0$
    & $0.40_{\pm0.30}$ {\scriptsize(5/5)}
    & $2.92_{\pm2.39}$ {\scriptsize(5/5)}
    & $\mathbf{0.56_{\pm0.30}}$ {\scriptsize(5/5)} \\
  & $t\;(\mathrm{s}) \in [2.0,\;3.0]$
    & $0.00_{\pm0.00}$ {\scriptsize(2/5)}
    & $0.32_{\pm0.32}$ {\scriptsize(5/5)}
    & $\mathbf{0.00_{\pm0.00}}$ {\scriptsize(5/5)} \\
\addlinespace
\multirow{2}{*}{S10}
  & $v\;(\mathrm{m/s}) \in [5.0,\;10.0]$
    & $0.00_{\pm0.00}$ {\scriptsize(5/5)}
    & ---
    & $\mathbf{0.00_{\pm0.00}}$ {\scriptsize(5/5)} \\
  & $t\;(\mathrm{s}) = 1.0$
    & ---
    & ---
    & $\mathbf{0.07_{\pm0.05}}$ {\scriptsize(3/5)} \\
\bottomrule
\end{tabular}%
}
\end{table}

\subsection{Ablation Study of Draft Construction}
\label{sm::AbDraft}

We ablate both draft construction components on 200 scene descriptions using GPT-4.1, comparing three configurations: (i)~\textbf{single-shot prompt}, which generates a complete ESG in one response with no retry or validation; (ii)~\textbf{w/o Validation}, which applies
hierarchical prompting but passes drafts directly downstream without
the validation step; and (iii)~\textbf{Ours}, which combines hierarchical prompting with full validation. Final usability and high-quality rate are evaluated as defined in Sec.~\ref{exp::AbDraft}; results are reported in Table~\ref{tab:ablation_draft}.

Hierarchical prompting improves usability by decomposing ESG
generation into progressively constrained stages, reducing
schema errors and reference inconsistencies that single-shot
outputs frequently produce.
The validation system further raises final usability to 99.0\%
by filtering malformed or contradictory drafts before they enter
the optimization pipeline, and raises the high-quality rate by
ensuring that only structurally sound drafts---those admitting
a valid low-energy layout---proceed downstream.

\subsection{Ablation Study of Gradient Strategy}
\label{sm::AbGrad}

\paragraph{Scene and setup.}
We use scene S6 (car $\to$ trashbin $\to$ table), which involves non-spherical collision proxies: the car is represented by a box primitive and the trashbin by a cylindrical proxy. This geometry makes the scene a suitable stress test for gradient quality: the first event (car $\to$ trashbin) is a direct interaction whose gradient is relatively straightforward, while the second event (trashbin $\to$ bench) is a chain interaction where the gradient must backpropagate through the non-spherical car--trashbin contact. We run each variant for $K{=}5$ independent runs with different random initializations of the car's initial speed.

\paragraph{Why AD fails on chain events with non-spherical geometry.} For sphere--sphere contacts, the contact normal is exactly center-to-center, so the AD surrogate gives accurate gradient directions. For non-spherical primitives (boxes, cylinders), the contact normal depends on which face or edge is closest---a quantity that the penalty-based soft-contact model approximates via the primitive SDF. This approximation introduces a directional error in the contact impulse, which propagates as gradient noise when backpropagating through the chain. In practice, the AD gradient for the first event remains reliable (5/5 completions), but the chain gradient for the second event becomes sufficiently inaccurate to prevent convergence (0/5 completions). FD gradients, evaluated on the full Newton simulator with exact contact resolution, avoid this issue at the cost of $2|\theta|$ additional forward passes per step.

\paragraph{Per-run results.}

\begin{center}\small
\begin{tabular}{lcrrrr}
\toprule
 & Run & Ev.\,1 & Ev.\,2 & Time\,(s) \\
\midrule
\multirow{5}{*}{AD-only}
 & 1 & \checkmark & \texttimes & 33 \\
 & 2 & \checkmark & \texttimes & 35 \\
 & 3 & \checkmark & \texttimes & 36 \\
 & 4 & \checkmark & \texttimes & 37 \\
 & 5 & \checkmark & \texttimes & 35 \\
\midrule
\multirow{5}{*}{FD-only}
 & 1 & \checkmark & \checkmark & 144 \\
 & 2 & \checkmark & \checkmark & 152 \\
 & 3 & \checkmark & \checkmark & 156 \\
 & 4 & \checkmark & \texttimes  & 155 \\
 & 5 & \checkmark & \checkmark & 160 \\
\midrule
\multirow{5}{*}{Hybrid (Ours)}
 & 1 & \checkmark & \checkmark & 99 \\
 & 2 & \checkmark & \checkmark & 91 \\
 & 3 & \checkmark & \checkmark & 103 \\
 & 4 & \checkmark & \checkmark  & 112 \\
 & 5 & \checkmark & \texttimes & 101 \\
\bottomrule
\end{tabular}
\end{center}

\paragraph{Switching criterion.}
The AD$\to$FD switch is determined statically at planning time by inspecting the ESG dependency graph: if the active actor for event $e_{k+1}$ appears as a \emph{passive} body in event $e_k$ (i.e., it receives rather than initiates the upstream contact), the optimizer flags the chain as FD-required for all gradient steps involving $e_{k+1}$. This requires no runtime detection and adds no overhead beyond the FD gradient calls themselves.

\subsection{Ablation Study of Collision Model}
\label{sm::AbSDF}

\paragraph{Scene and setup.}
We use scene S6 (car $\to$ trashbin $\to$ table), isolating event-2 (trashbin--table collision) as the evaluation target.
The car's initial speed is optimized by our full Hybrid method, so both variants start from the same post-event-1 trashbin state; only the collision model for the trashbin--table contact is varied. Each variant is run $K{=}10$ times.

\paragraph{Simple proxy vs.\ SDF refinement.}
The simple-proxy variant represents the car, trashbin and table by their bounding-box sphere approximations (box or sphere). While sufficient for detecting rough proximity, this coarse geometry underestimates the effective contact offset and misestimates the contact normal direction for cylindrical and rectangular bodies, causing the optimizer to steer the trashbin along an incorrect trajectory that bypasses the bench.
The SDF-refinement variant follows a two-stage procedure: it first runs the simple-proxy optimizer to obtain a warm-start parameter, then runs an additional refinement pass in which contact distances and normals are computed from voxel SDFs loaded for both actors. The higher-fidelity surface gradients provide a more accurate steering signal for the narrow-corridor approach required to make the trashbin contact the table.

\paragraph{Actual results.}

\begin{center}\small
\begin{tabular}{lcc}
\toprule
Method & Event-2 successes & Success rate \\
\midrule
Simple proxy only   & 1/10 & 10\% \\
Proxy + SDF (Ours)  & 8/10 & 80\% \\
\bottomrule
\end{tabular}
\end{center}

\noindent
The single success of the simple-proxy baseline occurs on a run
whose random initialization happened to place the trashbin on a
near-optimal trajectory, confirming that the baseline can only
succeed by chance rather than by reliable gradient guidance.
The 80\% success rate of SDF refinement reflects the residual
difficulty of the narrow-target geometry; the two failures arise
from initializations where the warm-start proxy solution is too
far from the SDF solution basin to be recovered within 30 steps.

\subsection{More Ablation Studies}
\label{sm::MoreAb}
\begin{figure}
    \centering
    \includegraphics[width=1\linewidth]{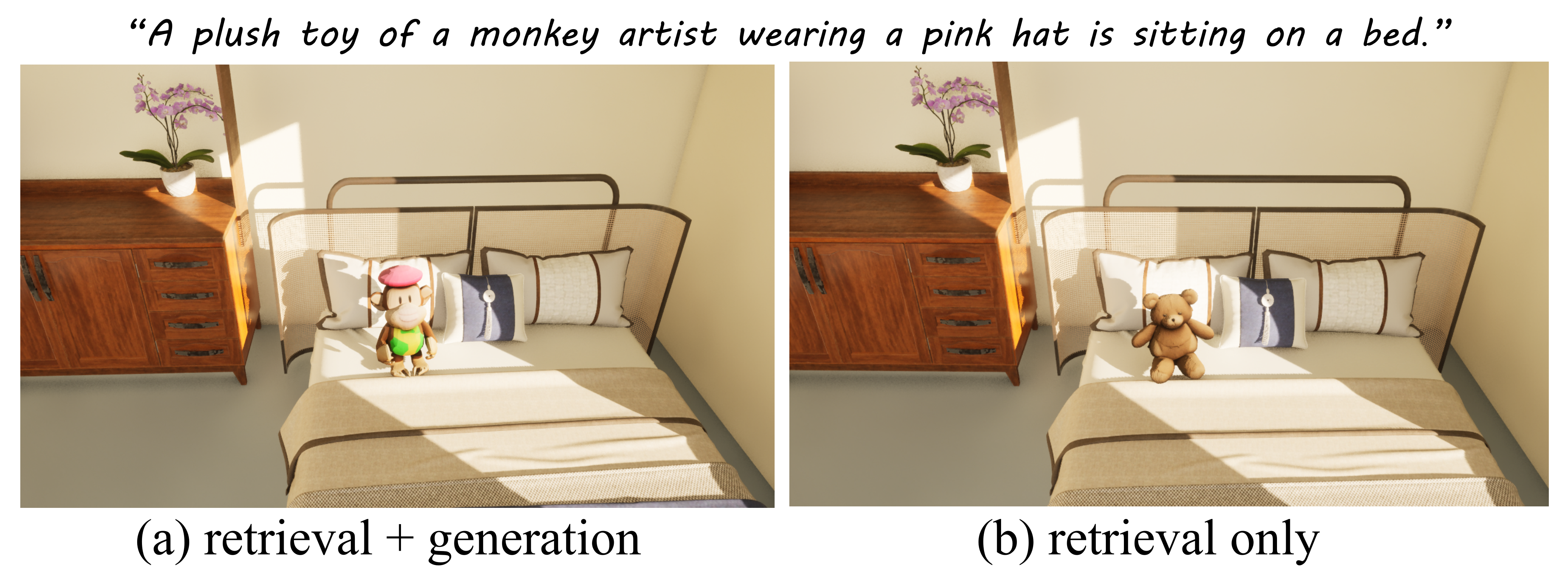}
    \caption{Ablation study of Asset Generation. Retrieval-only method is limited by the capacity of asset library, forcing to return the matched object(a teddy bear in this case), while asset generation module can better meet the user's unique requirements.}
    \label{experiment::ablation4}
\end{figure}

\paragraph{Ablation Study of Asset Generation}
\label{sm::AbAsset}
We ablate our asset generation module, which conditionally invokes Trellis to synthesize a 3D asset when library retrieval confidence falls below an empirically chosen threshold. We design prompts that include a user-specified object unlikely to be covered by the asset library and compare two variants under identical settings: (i) \textit{retrieval-only}, which directly substitutes the object with the closest library match, and (ii) \textit{retrieval + generation} (ours), which generates the missing asset from text and imports it before scene construction.
As shown in ~\refFig{experiment::ablation4}, enabling asset generation apparently produces assets that better match the requested object, allowing the pipeline to satisfy open-vocabulary user intents beyond the library’s coverage.

\paragraph{Ablation Study of Dynamic Optimization}
\label{sec:ablation_dynamic_optimization}

\begin{figure}
    \centering
    \includegraphics[width=1\linewidth]{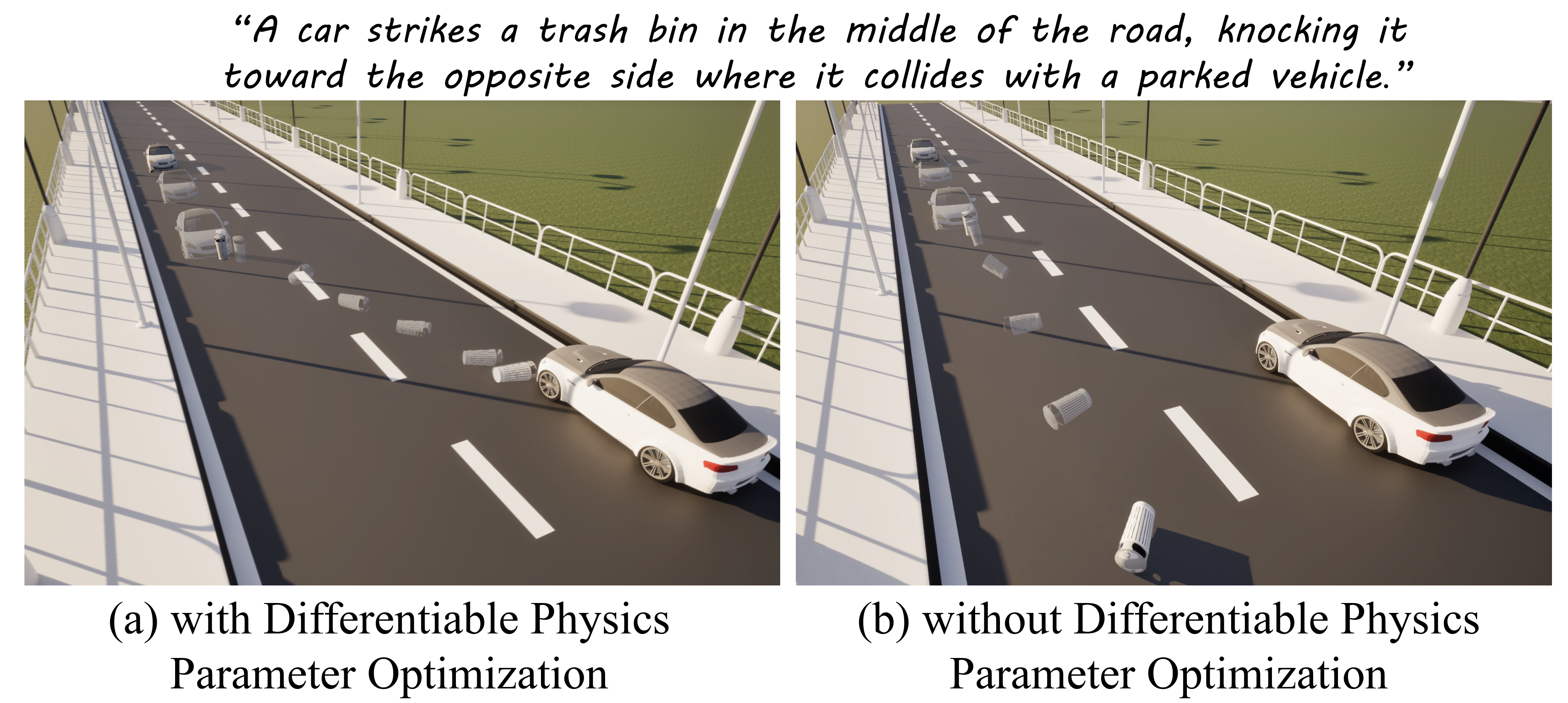}
    \caption{Ablation study of Differentiable physics optimization. For cascading multi-stage motion and long-sequence motion, initialization can hardly achieve the goal without differentiable physics optimization.}
    \label{experiment::Ablation3}
    \vspace{-0.3cm}
\end{figure}

Differentiable physics optimization is crucial for dynamic scene generation, since many event targets cannot be satisfied by directly assigning motion parameters with simple heuristics. We ablate this module by removing dynamic optimization and using only rule-based initialization of object parameters (e.g., coarse velocity estimates), while keeping the rest of the simulation pipeline identical.
As shown in ~\refFig{experiment::Ablation3}, in a cascading scenario,  the initialization-only baseline can often trigger the first contact (car → bin) but cannot reliably enforce the downstream event (bin → car). Only with dynamic optimization enabled, can the system adjust motion parameters to satisfy such multi-stage cascades as well as longer, more complex sequences.

\subsection{Human Evaluation on Ablation Study}
\label{sm::human}
We conduct a user study on 20 generated dynamic scenes to assess the perceptual quality. Each scene is evaluated by 37 experienced users from two perspectives: \emph{(i) scene quality}, focusing on physical plausibility and overall realism of motions and interactions, and \emph{(ii) input alignment}, measuring how well the generated scene adheres to the given textual specification. Both criteria are rated on a 5-point scale (1: very poor, 5: excellent). We compare the full system against four ablated variants: removing \emph{global optimization}, \emph{initial placement}, \emph{dynamic optimization}, and \emph{asset generation}. The result is shown in ~\refTab{tab:user_study}.

\begin{table}[!htbp]
\centering
\caption{User study on ablated variants, rated on a 5-point scale.}
\label{tab:user_study}
\resizebox{\columnwidth}{!}{
\setlength{\tabcolsep}{6pt}
\begin{tabular}{lcc}
\toprule
\textbf{Method} & \textbf{Scene quality} $\uparrow$ & \textbf{Input alignment} $\uparrow$ \\
\midrule
Ours (Full) & 4.4 $\pm$ 0.4 & 4.5 $\pm$ 0.4 \\
w/o Global optimization & 4.0 $\pm$ 0.5 & 2.7 $\pm$ 0.6 \\
w/o Initial placement & 3.8 $\pm$ 0.5 & 2.8 $\pm$ 0.9 \\
w/o Dynamic optimization & 4.4 $\pm$ 0.4 & 2.8 $\pm$ 0.8 \\
w/o Asset generation & 3.9 $\pm$ 0.5 & 2.3 $\pm$ 0.6 \\
\bottomrule
\end{tabular}
}
\end{table}

\subsection{Failure Case Analysis}

\begin{figure}
    \centering
    \includegraphics[width=1\linewidth]{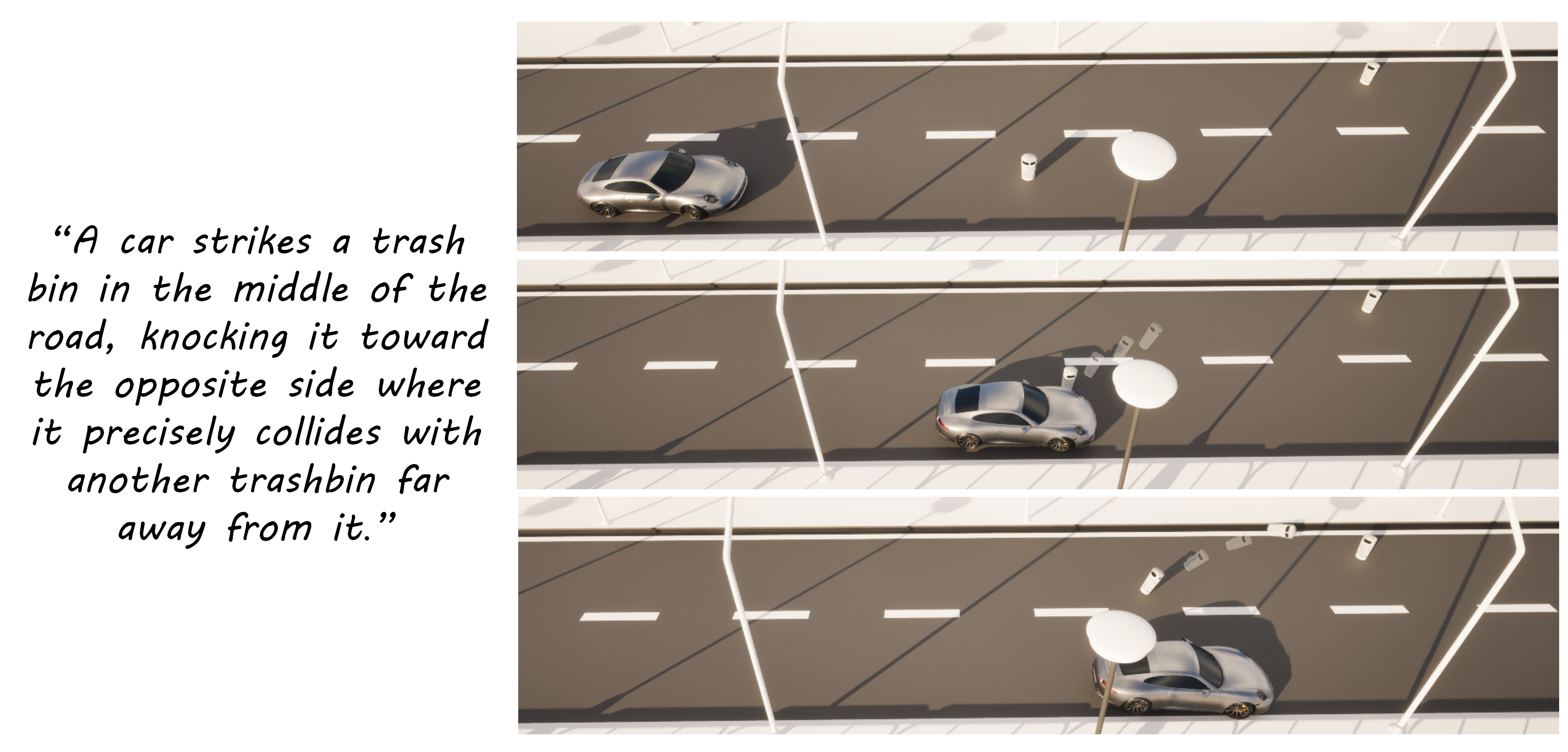}
    \caption{A typical failure case. User intends to let the trashbin collide with another trashbin, but the actual optimization sometimes fails to meet the precise demand due to the difference of proxy geometry.}
    \label{sm::failure}
\end{figure}

\paragraph{Complex model cascading events.}
In ~\refFig{sm::failure}, we examine a demanding multi-stage
scenario in which a car strikes a trash bin, which then travels to
precisely collide with a second bin placed farther away.
When such cascading requirements are specified with particularly tight spatial or temporal tolerances, the success rate can decrease compared to simpler scenes.
This stems from a gap between the differentiable collision proxies
used during optimization and the true contact geometry executed by
Unreal Engine at runtime: for complex assets such as vehicles and
bins, small differences in contact offsets and post-impact
trajectories can accumulate across stages, occasionally causing the
downstream collision to be missed in the UE replay even when the
proxy-based rollout satisfies the objective.
Relaxing the event tolerance or subdividing the cascade into
separately optimized stages largely mitigates this effect.

\paragraph{Physically unreasonable prompts.}
Our system prioritizes physical consistency over literal compliance
when a prompt specifies behaviors irreconcilable with rigid-body
dynamics.
As an illustrative example, consider the prompt:
\emph{``A light ball moving at speed $v$ strikes a heavy stationary
ball; the heavy ball moves away at speed $v$ after the collision with no other forces.''}
This violates momentum conservation: a lighter impactor cannot
transfer its full speed to a heavier target, so the prescribed
post-collision velocity is physically unattainable.
Rather than forcing the scene to satisfy the literal requirement,
the optimizer seeks the nearest feasible configuration---maximizing
the heavy ball's post-collision speed within the bounds allowed by
the mass ratio and impact geometry---while the physical prior
$\mathcal{L}_\text{prior}$ prevents parameters from drifting to
non-physical values.
The resulting scene is executable in the physics engine and as close
to the user's intent as physics permits, though the heavy ball
reaches a lower speed than specified.
This behavior reflects an intended design principle: the rigid-body
simulation loop acts as a feasibility filter, ensuring that generated
scenes remain grounded in real dynamics even when user intent is
over-specified or physically contradictory.

\section{\esg~ Schema Specification}
\label{app:esg_schema}

We adopt a unified, structured scene representation termed \esg~(\nn) to serve as a machine-checkable and reading-friendly contract across all modules. Specifically, \nn~ is a JSON object whose top-level fields are organized into four core components shown in the canonical structure below,
\begin{lstlisting}[language=json,caption={Canonical ESG skeleton.},label={lst:esg_skeleton}]
{
  "scene_definition": { ... }, 
  "asset_requirement": { ... }, 
  "spatial_relation_graph": { ... },  
  "timeline_system": { ... }
}
\end{lstlisting}
In practice, ESG is refined through four sequential stages that mirror the modular structure of the pipeline described in Section 3. The first stage, detailed in Section 3.1 under Layered Draft Construction, translates the user prompt into a schema conformant ESG draft by instantiating object intents, coarse spatial relations, and an initial sketch of the event timeline. The second stage, presented in Section 3.2 as Asset Resolution, grounds these semantic drafts by resolving each \emph{asset\_requirement} into concrete, engine-ready resources through retrieval or on-demand generation, while attaching canonical geometry and physics profiles to establish deterministic asset bindings. The third stage, introduced in Section 3.3 as Static Layout Solving, converts the semantic relation graph into an explicit static scene layout by jointly solving object poses and scales under physical feasibility, while respecting all specified structural constraints and semantic regions. The final stage, described in Section 3.4 as Dynamic Optimization, completes the temporal program by optimizing event timings and motion-related physical parameters under simulation constraints, producing an executable timeline composed of atomic control actions that can be directly replayed in the engine.


\subsection{scene\_definition: Metadata of Scene Description}

\begin{lstlisting}[language=json,caption={brief example of scene\_definition.},label={lst:scene_definition}]
"scene_definition":{
  "scene_id": { ... }, 
  "coordinate_system": { ... }, 
  "original_description": { ... },  
  "expanded_description": { ... },
  "scene_parameters": { 
    "is_dynamic": ...,
    "scene_type": ...,
    ...
  }
}
\end{lstlisting}

The unique scene\_id is for identifying different scene files. The coordinate\_system block specifies the coordinate convention, units, and world bounds. The paired text fields original\_description and expanded\_description that retain the user input and the LLM-expanded narrative used by downstream modules. Finally, more scene\_parameters such as is\_dynamic and scene\_type are also stored in this region.

\subsection{Asset Requirement \& Configuration}
\label{app:esg_asset_layer}

ESG represents assets using a two-level abstraction to explicitly decouple \emph{what is needed} from \emph{what is used}.

\emph{asset\_requirement} is a dictionary keyed by unique asset\_id given by LLM. Each value specifies the desired category (\emph{asset\_category}), descriptive attributes (\emph{asset\_description}), and coarse physical priors (e.g., \emph{expected\_dimensions\_cm}, \emph{expected\_mass\_kg}, \emph{is\_movable}). These fields are produced in stage0 and are used for asset retrieval(or generation), layout reasoning and physical initialization.
After retrieval(or generation), this region becomes \emph{asset\_configurations} in stage1, binding each semantic requirement to a concrete engine asset, including identifiers and file paths, realized scale, orientation, and physics-relevant attributes. 

\subsection{Spatial Relation Graph: semantic description of scene layout}

Static structure is expressed as a directed relation graph \(G\) including following fields:

\begin{itemize}
  \item \textbf{Semantic regions} defines named placement zones. Every object must belong to a semantic region, and every semantic region is provided by its \emph{parent object}. For example, wooden\_table provides a table\_face region, so wooden table is parent object of table\_surface.
  \item \textbf{Nodes.} Each node is corresponding to an object in the scene, having a unique node\_id. Each node includes an \texttt{asset\_id} that must refer to a key in \texttt{asset\_requirement}, which means it is an instance of the referred asset. The node should also make clear of its semantic region, or announce to be a structural object if it make up the basic structure of the scene(e.g. ground, wall)
  \item \textbf{Edges.} Relations includes binary\_relations and multi\_relations, each represented by a directed edge with specific \emph{relation\_type}, confidence score and a brief description of this relation.
\end{itemize}

Common relations include "on, under, in\_front\_of, behind, left\_of, right\_of, near, adjacent, far\_from, above, higher\_than, lower\_than, same\_level\_as, face, inside\_of, center\_of, attached\_vertical, \\attached\_horizontal, between, surround, align, scatter\_by".

The following part is a brief example of the fields of spatial relation graph, with some values omitted.
\begin{lstlisting}[language=json,caption={brief example of spatial\_relation\_graph.},label={lst:spatial_relation_graph}]
"spatial_relation_graph": {
    "semantic_regions": {
      "floor_area_1": {
        "region_description": "...",
        "region_type": "SUPPORT_SURFACE",
        "parent_object": "floor_1"
      },
      ...
    }
    "nodes": {
      "table_1": {
        "asset_id": "modern_office_desk_table_01",
        "semantic_region": "floor_area_1"
      },
      ...
    }
    "edges": {
      "binary_relations": [
        {
          "from": "desk_1",
          "to": "floor_1",
          "relation_type": "on",
          "confidence": 0.99,
          "semantic_description": "..."
        },
        ...
      ]
    }
}
\end{lstlisting}

\subsection{Timeline System: sequential schedule of dynamic process}
Only required if the scene is dynamic. ESG encodes both \emph{what should happen} and \emph{what is physically controllable}.

\paragraph{Actors and controllability.}
\emph{actors} assigns each object instance a role (active/passive). Active objects can move actively, whose parameters can be controlled. While passive objects can only be forced to move, with their parameters unable to adjust. Active actors should further specify a \emph{force\_direction} constraint that restricts the directions in which controlling forces can be applied, with allowed values including ANY\_3D (unconstrained), HORIZONTAL\_PLANE (forces confined to the XY plane), and ALONG\_FORWARD\_OR\_BACKWARD (forces aligned with the object's forward axis in either direction).

\paragraph{Event sequence graph.}
Temporal structure is encoded as a sequence graph consisting of event nodes and event edges. Nodes represent milestones (e.g., SceneStart, Collision, PositionAt) with timestamps or time windows; edges represent high-level continuous actions (e.g., UniformMotion, Toward) linking milestones and carrying physical parameters or their ranges. The duration of each action is defined by the time difference between its source and target milestone events. All parameters—including event timestamps and action-related physical quantities—can be specified either as fixed values or as bounded intervals, depending on the user input. Fixed-value parameters are treated as hard constraints and remain not optimizable, while interval-bounded parameters become optimization variables during the dynamic solving stage.

The following part is a brief example of the fields of timeline\_system, with some values omitted.
\begin{lstlisting}[language=json,caption={brief example of timeline\_system.},label={lst:timeline}]
"timeline_system": {
    "actors": {
      "car_1": {
        "role": "active",
        "force_direction": "ALONG_FORWARD_OR_BACKWARD"
      },
    },
    "sequence_graph": {
      "nodes": [
        {
          "id": "scene_start",
          "event": "SceneStart",
          "timestamp": 0.0
        },
        {
          "id": "car_trashbin_collision",
          "event": "Collision",
          "participants": [
            "car_1",
            "trashbin_1"
          ],
          "window": [1.7,1.9],
          "constraint": {...}
        }
      ]
      "edges": [
        {
          "id": "car_uniform_motion_before_collision",
          "from": "scene_start",
          "to": "car_trashbin_collision",
          "event": "UniformMotion",
          "actor": "car_1",
          "target": "trashbin_1",
          "params": {
            "speed": [100,3000],
            ...
          }
        }
      ]
    }
}
\end{lstlisting}

\section{System Prompts}
Our prompt system is organized into multiple layers; due to space limitations, we only include a subset of representative prompt fragments that capture the key analysis steps.

\subsection{Asset Requirement Analysis Prompt}

\begin{promptbox}{Scene Asset Expansion \& Specification Prompt}

Analyze the user’s scene description and plan the object list through the following steps:

\begin{enumerate}[label=\arabic*), wide=0pt]

\item Include all user-mentioned objects (mandatory).

Every object explicitly mentioned by the user must appear as an object instance in the scene. If the user mentions an object that does not exist in the predefined category list, map it to the closest available category rather than inventing a new one.

\item Decide whether supplementary objects are allowed.

\begin{itemize}

  \item If the user explicitly restricts the scene (e.g., ``only A'', ``no other objects''), do not add supplementary objects beyond what the user mentioned (plus necessary structural/base elements).

  \item Otherwise, perform controlled expansion: add only objects that logically belong to the scene type and support the user’s intent.

\end{itemize}

\item Perform controlled expansion (when allowed).

Add supplementary objects conservatively, guided by the following principles:

\begin{itemize}

  \item Logical necessity: add only items that naturally co-occur (e.g., a table typically implies a chair).

  \item Scene appropriateness: choose props that fit the described environment (e.g., study scenes: books/lamp; street scenes: street elements).

  \item Avoid overcrowding: do not turn a simple request into a cluttered scene; added objects must remain secondary to user-mentioned objects.

\end{itemize}

\item Name objects logically and avoid duplication.

Use only logical instance names (e.g., desk\_1, laptop\_1, room\_1). Never use asset-style identifiers as object names.

Ensure one clear object instance per real object---do not create duplicates by mixing ``asset identifiers'' and ``object instances'' as if they were separate physical objects.

\item Specify assets with matching-oriented descriptions.

For every required asset, write a rich description aimed at retrieval/matching. Prioritize information in this order:

\begin{enumerate}[label=\arabic*)]

\item Key features (e.g., with drawers, cushioned)

\item Material, color and style

\item Purpose/function

\item Typical usage scenario

\end{enumerate}

\item Assign plausible physical properties.

Provide reasonable dimensions (in centimeters) and mass(in kg) for each asset, consider its daily properties and usage. Use sensible linear damping: keep a default damping 0.1 unless the user description clearly implies unusual damping. Treat structural/base elements as non-movable; most other objects should be movable.

\item Ensure coverage and consistency.

\begin{itemize}

  \item Each requested asset should be referenced by at least one object instance.

  \item Multiple object instances may share the same asset when appropriate.

  \item Avoid redundant ``placeholder'' entries that do not correspond to an actual object instance.

\end{itemize}

\end{enumerate}

\end{promptbox}

\subsection{Static Layout Analysis Prompt}

\begin{promptbox}{Spatial Relation Graph (Static Layout) Analysis Prompt}

Analyze the user’s scene description and given object list, and construct a static spatial relation graph in a structured, step-by-step manner. Your goal is to produce a layout description that is metrically plausible, semantically consistent, and constraint-ready for downstream placement/optimization.

\begin{enumerate}[label=\arabic*), wide=0pt]
\item Determine the scene context and scope
\begin{itemize}
  \item Decide whether the scene is indoor or outdoor.
  \item Identify foreground actors vs.\ background context, keeping the scene complete but not overcrowded.
\end{itemize}
\item Establish the structural base (scene frame)
\begin{itemize}
  \item Indoor: define a minimal enclosing room frame (walls/ceiling/floor) so the space is a physically meaningful container.
  \item Outdoor: define exactly one ground/road base representing the entire walkable/drivable surface; do not split it into multiple ground-like bases.
\end{itemize}
\item Create semantic regions (functional placement domains)

\begin{itemize}
  \item Define semantic regions as functional sub-spaces that objects can be placed on, in, or attached to, each grounded in a valid parent object.
  \item Support surfaces: for objects that can support others (e.g., floor, table/desk tops, benches, road surface).
  \item Storage regions: for objects that can contain others (e.g., cabinet interior, car interior, boxes).
  \item Attached surfaces: vertical (e.g., wall surface) and horizontal (e.g., ceiling surface) attachment domains.
  \item Ensure regions are sufficient for all non-structural objects; avoid regions that do not correspond to any real parent object.
\end{itemize}
\item Assign each object to exactly one semantic region
\begin{itemize}
  \item Every regular (non-structural) object must belong to a meaningful semantic region consistent with its physical placement.
  \item Do not duplicate region membership as relation edges; region assignment already encodes that constraint.
\end{itemize}
\item Define object nodes (instances) consistently
\begin{itemize}
  \item Use logical object instance names (e.g., desk\_1, chair\_1, car\_1) and keep them consistent across regions and relations.
  \item Keep regular objects compact; if too many are mentioned, prioritize the most important ones to stay within a practical complexity budget.
\end{itemize}
\item Add spatial relations as canonical edges (only when meaningful)

\begin{itemize}
  \item Add relations only to express object-to-object semantics not already implied by semantic-region membership.
  \item Use canonical relation types (binary and multi-object) from the allowed inventory; map informal wording (e.g., ``next to'', ``against'') to the closest canonical relation (e.g., near or adjacent).
  \item Prefer relations within the same semantic region; avoid cross-domain relations unless strongly required by the description.
  \item Write relation descriptions unambiguously and check for contradictions, especially when multiple relations connect the same objects.
\end{itemize}
\item Enforce a single ``face'' relation per regular object

\begin{itemize}
  \item For every non-structural object, assign exactly one directional face relation representing its functional/usage direction in the horizontal plane.
  \item Strategy: indoor tables/desks face a fixed cardinal direction (e.g., +Y); outdoor road vehicles face along the road (+Y or -Y); most other objects should face a relevant object (e.g., chair faces desk/table) rather than a cardinal direction.
\end{itemize}

\item Final self-check (consistency and completeness)

\begin{itemize}
  \item Structural base matches scene type and is not duplicated.
  \item Every semantic region has a valid parent object and clear function.
  \item Every regular object is assigned to a valid semantic region.
  \item Every regular object has exactly one outgoing face relation.
  \item All relation types are canonical; no internal conflicts.
  \item The graph is not over-specified (avoid redundant constraints already implied by region assignment).
\end{itemize}
\end{enumerate}

\end{promptbox}

\subsection{Dynamic Timeline Analysis Prompt}

\begin{promptbox}{TIMELINE SYSTEM (DYNAMIC PARSING) PROMPT}

Analyze the user’s description and derive a timeline-constrained dynamic specification as a sequence graph (point events + segment motions). Focus on what must happen and when, while leaving continuous motion details to physics-based optimization.

\begin{enumerate}[label=\arabic*), wide=0pt]

\item Decide whether to generate a dynamic timeline
Only generate the dynamic timeline when the scene is intended to be dynamic. If the description is purely static, do not introduce a timeline.

\item Identify all actors and classify roles from the user description
List every object that is expected to move or be affected by motion, including objects that may be pushed, hit, or displaced by others.

Classify each listed actor based on the user description (not on object type):
\begin{itemize}
  \item active: the user describes it as moving by its own power or intention (controllable motion segments such as driving, braking, accelerating, turning)
  \item passive: the user describes it as stationary or only moved by external forces (e.g., being hit, pushed, blown)
\end{itemize}

If an active actor exists, assign a force-direction constraint appropriate to the described motion (e.g., any 3D, horizontal plane, along forward/backward, along forward only).

\item Construct point events (nodes) as event-level constraints
Use point events to represent sparse supervision that the physics should satisfy.
Include the following types when needed:
\begin{itemize}
  \item SceneStart: mandatory start event at time 0.0
  \item Collision: used when the user explicitly describes a collision between two objects
  \item Trigger: used to mark the start or end of an intentional controlled action by an active actor (e.g., start braking, stop accelerating)
  \item PositionAt: only use when the user explicitly requires a position constraint at a certain time, or when a minimal verification point is necessary to disambiguate a motion outcome
\end{itemize}

Only include the objects directly involved in a point event. Do not create trigger events for passive objects.

\item Plan time windows by building and validating a dependency graph
Derive a directed acyclic dependency graph from the described sequence (what happens before what).
Assign timestamps only when the user gives explicit times; otherwise assign time windows.

Plan windows in a topological order and enforce temporal consistency:
for every connection from node A to node B, ensure B starts after A can possibly end (the earliest time of B is not earlier than the latest time of A).

When the user does not specify times:
\begin{itemize}
  \item infer reasonable windows based on scene scale and typical physics
  \item keep windows relatively broad for uncertain events
  \item leave explicit gaps between sequential events to avoid overlap
\end{itemize}

After assignment, validate all dependencies again to ensure global consistency, especially for chain events.

\item Derive segment motions (edges) as abstract actions between point events
Represent continuous motion as segments that connect two point events.
Each segment describes the motion of one active actor from its start node to its end node. The segment duration is implied by its endpoints unless the user provides a specific duration.

Use segment types such as:
\begin{itemize}
  \item UniformMotion (steady motion)
  \item Brake (deceleration represented via increased damping or an equivalent braking control)
  \item AcceleratedMotion (speeding up)
  \item Turn (heading change)
\end{itemize}

Do not over-author post-event motion that the user did not describe (e.g., do not script how objects settle after a collision unless the user explicitly describes additional motion after the collision).

\item Handle multi-stage cascading motion with virtual inertial connectors when required
Multi-stage motion (e.g., A hits B, then B hits C) should be parsed and represented explicitly when the user describes it.

Use a virtual inertial connector only when all conditions are met:
\begin{itemize}
  \item a passive object is involved
  \item the passive object is first moved by an upstream collision
  \item the user explicitly describes that it subsequently collides with another object
\end{itemize}

Do not use virtual connectors for:
single collisions with no described follow-up, active actor movements, or general graph connectivity.

\item Choose parameters from the user description and provide feasible ranges when unspecified
When the user provides exact values or ranges, follow them.
When the user does not specify values, provide wide but reasonable numeric ranges based on common sense and daily experience.

Do not use unknown placeholders for parameters.

\item Map language cues to timeline elements
Translate common expressions into event and segment choices:
\begin{itemize}
  \item hits, crashes into: collision point event
  \item starts braking, stops, begins to brake: trigger point event plus a braking segment
  \item drives, moves forward: uniform motion segment
  \item accelerates: accelerated motion segment
  \item turns left/right: turning segment
\end{itemize}

\item Final self-check for dynamic consistency
Before output:
\begin{itemize}
  \item every actor exists in the static spatial relation graph
  \item exactly one start event exists at time 0.0
  \item all point events and segments form a valid directed acyclic graph
  \item all time windows satisfy the dependency ordering constraints
  \item triggers are only used for active actors’ controlled actions
  \item virtual inertial connectors are only used for user-described chain collisions of passive objects
\end{itemize}

\end{enumerate}

\end{promptbox}


\end{document}